\documentclass{article}
\usepackage{amssymb}

\usepackage{amsmath}

\newtheorem{theorem}{Theorem}
\newtheorem{lemma}[theorem]{Lemma}
\newtheorem{remark}[theorem]{Remark}

\def\proof{\noindent{\bfseries Proof. }}
\def\endproof{\mbox{\ \rule{.1in}{.1in}}}

\begin{document}

\title{Sine-Gordon breather form factors\\
and quantum field equations}
\author{H. Babujian\thanks{%
Permanent address: Yerevan Physics Institute, Alikhanian Brothers 2,
Yerevan, 375036 Armenia.} \thanks{
e-mail: babujian@lx2.yerphi.am, babujian@physik.fu-berlin.de}~ and M.
Karowski\thanks{
e-mail: karowski@physik.fu-berlin.de} \\
Institut f\"ur Theoretische Physik\\
Freie Universit\"at Berlin,\\
Arnimallee 14, 14195 Berlin, Germany\\
}
\maketitle

\begin{abstract}
Using the results of previous investigations on sine-Gordon form factors
exact expressions of all breather matrix elements are obtained for several
operators: all powers of the fundamental bose field, general exponentials of
it, the energy momentum tensor and all higher currents. Formulae for the
asymptotic behavior of bosonic form factors are presented which are
motivated by Weinberg's power counting theorem in perturbation theory. It is
found that the quantum sine-Gordon field equation holds and an exact
relation between the ``bare'' mass and the renormalized mass is obtained.
Also a quantum version of a classical relation for the trace of the energy
momentum is proven. The eigenvalue problem for all higher conserved charges
is solved. All results are compared with perturbative Feynman graph
expansions and full agreement is found.\\[8pt]
\end{abstract}


\section{Introduction}

This work continuities previous investigations \cite{BFKZ,BK} on exact form
factors for the sine-Gordon alias the massive Thirring model. Some results
of the present paper have been published previously \cite{BK1}. The
classical sine-Gordon model is given by the wave equation 
\[
\square \varphi (t,x)+\frac{\alpha }{\beta }\sin \beta \varphi (t,x)=0. 
\]
Since Coleman \cite{Co} found the wonderful duality between the quantum
sine-Gordon and the massive Thirring model a great deal of effort have been
made to understand this quantum field theoretic model. The present article
is a further contribution in this direction. The main new results are:

\begin{enumerate}
\item  Using Weinberg's power counting theorem we prove in perturbation
theory that matrix elements of exponentials of a bose field satisfy a
'cluster property' in momentum space. We use this as a characterizing
property for exponentials of bose fields

\item  In \cite{BK} we introduced the concept of 'p-functions' which belong
to local operators for the sine-Gordon solitons (see also \cite{Sm3,Ta}).
Here we formulate that concept for the sine-Gordon breathers.

\item  We investigate the higher conservation laws which are typical for
integrable quantum field theories. Hereby we correct a mistake in the
literature (see the footnote \ref{c2}).

\item  We prove the quantum field equation of motion. Thus we derive
independently the matrix elements of the operators $\varphi $ and $:\!\sin
\beta \varphi \!:$ and show that they satisfy the field equation (after a
finite mass renormalization). Hereby we again correct some mistakes in the
literature (see the footnote \ref{c1}).
\end{enumerate}

In addition we also recall some known formulae in order to present a more
complete picture of the sine-Gordon breather form factors. The sine-Gordon
model alias the massive Thirring model describes the interaction of several
types of particles: solitons, anti-solitons alias fermions and anti-fermions
and a finite number of charge-less breathers, which may be considered as
bound states of solitons and anti-solitons. Integrability of the model
implies that the n-particle S-matrix factorizes into two particle S-matrices.

The ``bootstrap'' program for integrable quantum field theoretical models in
1+1 dimensions starts as the first step with the calculation of the
S-matrix. Here (see e.g. \cite{KTTW,K2}) our starting point is the two
particle sine-Gordon S-matrix for the scattering of fundamental bosons
(lowest breathers) \cite{KT}\footnote{%
This S-matrix element has been discussed before in \cite{VG,AK}.} 
\[
S(\theta )=\dfrac{\sinh \theta +i\sin \pi \nu }{\sinh \theta -i\sin \pi \nu }%
\,. 
\]
The pole of $S(\theta )$ at $\theta =i\pi \nu $ belongs to the second
breather $b_{2}$ as a breather-breather bound state. The parameter $\nu $ is
related to the sine-Gordon and the massive Thirring model coupling constant
by 
\[
\nu =\frac{\beta ^{2}}{8\pi -\beta ^{2}}=\frac{\pi }{\pi +2g} 
\]
where the second equation is due to Coleman \cite{Co}.

As a second step of the ``bootstrap'' program off-shell quantities as
arbitrary matrix elements of local operators 
\[
^{out}\left\langle \,p_{m}^{\prime },\ldots ,p_{1}^{\prime }\left| \mathcal{O%
}(x)\right| p_{1},\ldots ,p_{n}\,\right\rangle ^{in}\, 
\]
are obtained by means of the ``form factor program'' from the S-matrix as an
input. Form factors for integrable model in 1+1 dimensions were first
investigated by Vergeles and Gryanik \cite{VG} for the sinh-Gordon model and
by Weisz \cite{W}\footnote{%
Similar results were obtained by Zamolodchikov \cite{Za1}.} for the
sine-Gordon model. The `form factor program' was formulated in \cite{KW,BKW}
where the concept of generalized form factors was introduced. In that
article consistency equations were formulated which are expected to be
satisfied by these objects. Thereafter this approach was developed further
and studied in the context of several explicit models by Smirnov \cite{Sm}
who proposed the form factor equations (see below) as extensions of similar
formulae in the original article \cite{KW}. Further publications on form
factors and in particular on sine-Gordon and sinh-Gordon form factors are 
\cite{Sm1}--\cite{CF}. Smirnov's approach in \cite{Sm1} is similar to the
one used in the present article (see section \ref{s6}). Also there is a nice
application \cite{GNT,CET} of form factors in condensed matter physics for
example for the one dimensional Mott insulators.

Let $\mathcal{O}(x)$ be a local operator. The generalized form factors $%
\mathcal{O}_{n}(\theta _{1},\dots ,\theta _{n})$ are defined by the vacuum
-- n-lowest breather matrix elements 
\[
\langle \,0\,|\,\mathcal{O}(x)\,|\,p_{1},\dots ,p_{n}\,\rangle
^{in}=e^{-ix(p_{1}+\dots +p_{n})}\,\mathcal{O}_{n}(\theta _{1},\dots ,\theta
_{n})~,~~~\mathrm{for}~~\theta _{1}>\dots >\theta _{n} 
\]
where the $\theta _{i}$ are the rapidities of the particles $p_{i}^{\mu
}=m(\cosh \theta _{i},\sinh \theta _{i})$. In the other sectors of the
variables the functions $\,\mathcal{O}_{n}(\theta _{1},\dots ,\theta _{n})$
are given by analytic continuation with respect to the $\theta _{i}$.
General matrix elements are obtained from $\mathcal{O}_{n}(\underline{\theta 
})$ by crossing which means in particular the analytic continuation $\theta
_{i}\rightarrow \theta _{i}\pm i\pi $.

In \cite{KW} one of the present authors (M.K.) and Weisz showed that for the
case of a diagonal S-matrix the n-particle form factor may be written as 
\begin{equation}
\mathcal{O}_{n}(\underline{\theta })=K_{n}^{\mathcal{O}}(\underline{\theta }%
)\prod_{1\leq i<j\leq n}F(\theta _{ij})  \label{1.1}
\end{equation}
where $\theta _{ij}=|\theta _{i}-\theta _{j}|$ and $F(\theta )$ is the
two-particle form factor (see section \ref{s2}). The K-function is an even $%
2\pi i$ periodic meromorphic function. In \cite{BK} we presented a general
formula for soliton anti-soliton form factors in terms of an integral
representation. Using the bound state fusion method we derived the general
soliton-breather and pure breather form factor formula which we will
investigate in this article in more detail. In particular for the case of
lowest breathers the K-function turns out to be of the form\footnote{\label%
{f2}For the sinh-Gordon model and the case of the exponential field
Brazhnikov and Lukyanov \cite{BL} found by different methods a formula which
agrees with our results. Smirnov \cite{Sm1} derived an integral
repesentation of sine-Gordon breather form factors (see subsection \ref{s6.2}%
) which agrees for some cases with our results (e.g. for the current and the
energy momentum tensor) and in the case of the higher currents not (see also
footnote \ref{c2}).} 
\begin{equation}
K_{n}^{\mathcal{O}}(\underline{\theta })=\sum_{l_{1}=0}^{1}\dots
\sum_{l_{n}=0}^{1}(-1)^{l_{1}+\dots +l_{n}}\prod_{1\leq i<j\leq n}\left(
1+(l_{i}-l_{j})\frac{i\sin \pi \nu }{\sinh \theta _{ij}}\right) p_{n}^{%
\mathcal{O}}(\underline{\theta },\underline{l})\,.  \label{1.2}
\end{equation}
The breather p-function $p_{n}^{\mathcal{O}}(\theta _{1},\dots ,\theta
_{n},l_{1},\dots ,l_{n})$ encodes the dependence on the operator $\mathcal{O}%
(x)$. It is obtained from the solitonic p-function $p_{sol,2n}^{\mathcal{O}}(%
\tilde{\theta}_{1},\dots ,\tilde{\theta}_{2n};\allowbreak z_{1},\dots ,z_{n})
$ (see \cite{BK}) by setting $\tilde{\theta}_{2i-1}=\theta _{i}+\tfrac{1}{2}%
iu^{(1)}\,,\;\tilde{\theta}_{2i}=\theta _{i}-\tfrac{1}{2}iu^{(1)}$ and $%
z_{j}=\theta _{j}-\tfrac{i\pi }{2}(1-(-1)^{l_{j}}\nu )$ with the fusion
angle $u^{(1)}=\pi (1-\nu )$. In \cite{BK} we proposed the solitonic
p-functions for several local operators. In this way we obtained all
breather p-functions which are some sort of decedents of solitonic
p-functions i.e. we just used the solitonic ones in the bound state points.
In the present article we will take a somewhat different point of view. We
will obtain a wider class of p-functions corresponding to local operators
with respect to the breather field, including also operators which are
non-local with respect to the solitonic field. The alternative point of view
is the following: As already mentioned, it has been shown in \cite{KW} that
a form factor of n fundamental bosons (lowest breathers) is of the form (\ref
{1.1}) where the K-function $K_{n}^{\mathcal{O}}(\underline{\theta })$ is
meromorphic, symmetric and periodic (under $\theta _{i}\rightarrow \theta
_{i}+2\pi i$). In addition it has to satisfy some additional conditions (see
section \ref{s2}). We consider eq.~(\ref{1.2}) as an Ansatz for the
K-function which transforms these conditions on the K-functions to simple
equations for the p-functions. In section \ref{s4} we present solutions of
these equations.

In this article we propose the p-functions for several local operators. In
particular we consider the infinite set of local currents $J_{L}^{\pm
}(x),\;\,(L=\pm 1,\pm 3,\dots )$ belonging to the infinite set of
conservation laws which are typical for integrable quantum field theories.
For this example the correspondences between the operators, the K-functions
and the p-functions are (up to normalization constants) 
\[
J_{L}^{\pm }(x)\leftrightarrow K_{n}^{(L,\pm )}(\underline{\theta }%
)\leftrightarrow p_{n}^{(L,\pm )}(\underline{\theta },\underline{l}%
)\varpropto \sum_{i=1}^{n}e^{\pm \theta _{i}}\sum_{i=1}^{n}e^{L\left( \theta
_{i}-\frac{i\pi }{2}(1-(-1)^{l_{i}}\nu )\right) }. 
\]
Here the breather p-function is obtained from the corresponding solitonic
one \cite{BK}.

In contrast to this case the breather p-function for exponentials $%
:\!e^{i\gamma \varphi }\!:\!(x)$ of the field $\varphi $ for generic real $%
\gamma $ is not related to a solitonic p-function of any local operator
(which means that $:\!e^{i\gamma \varphi }\!:\!(x)$ is not local with
respect to the soliton field). Now the correspondences are 
\[
:\!e^{i\gamma \varphi }\!:\!(x)\leftrightarrow K_{n}^{(q)}(\underline{\theta 
})\leftrightarrow p_{n}^{(q)}(\underline{l})=N_{n}^{(q)}%
\prod_{i=1}^{n}q^{(-1)^{l_{i}}} 
\]
where $q=q(\gamma )$ (see section \ref{s4}). Here and in the following $%
:\cdots :$ denotes normal ordering with respect to the physical vacuum. This
means in particular for the vacuum expectation value $\langle
\,0\,|\!:\!\varphi ^{N}\!:\!(x)\,|\,0\,\rangle =0$ and therefore $\langle
\,0\,|\!:\!\exp i\gamma \varphi \!:\!(x)|\,0\,\rangle =1$. In section \ref
{s4} we present arguments to support these correspondences and also
determine the normalization constants $N_{n}^{\mathcal{O}}$.

As an application of these results we investigate quantum operator
equations. In particular we provide exact expressions for all matrix
elements of all powers of the fundamental bose field $\varphi (t,x)$ and its
exponential $:\!\exp i\gamma \varphi \!:\!(t,x)$ for arbitrary $\gamma $. We
find that the operator $\square ^{-1}\!:\!\sin \beta \varphi (x)\!:$ is
local. Moreover the quantum sine-Gordon field equation\footnote{%
In the framework of constructive quantum field theory quantum field
equations where considered in \cite{Sch,F}. For the sine-Gordon model
quantum field equations where discussed by Smirnov in \cite{Sm1} and for the
sinh-Gordon model in \cite{MS} (see also footnote \ref{c1}).} 
\[
\square \varphi (x)+m^{2}\frac{\pi \nu }{\sin \pi \nu }\frac{1}{\beta }%
:\!\sin \beta \varphi \!:\!(x)=0 
\]
is fulfilled for all matrix elements. The factor $\frac{\pi \nu }{\sin \pi
\nu }$ modifies the classical equation and has to be considered as a quantum
correction of the breather mass $m$ as to be compared with the ``bare'' mass 
$\sqrt{\alpha }$. Further we find that the trace of the energy momentum
tensor $T^{\mu \nu }$ satisfies 
\[
T_{~\mu }^{\mu }(x)=-2\frac{\alpha }{\beta ^{2}}\left( 1-\frac{\beta ^{2}}{%
8\pi }\right) \left( :\!\cos \beta \varphi \!:\!(x)-1\right) . 
\]
Again this operator equation is modified by a quantum correction $(1-\beta
^{2}/8\pi )$ compared to the classical one.

Also we show that the higher local currents $J_{M}^{\mu }(t,x)$ satisfy $%
\partial _{\mu }J_{M}^{\mu }(t,x)=0$ and calculate all matrix elements of
all higher conserved $Q_{L}=\int dxJ_{L}^{0}(t,x)$%
\begin{equation}
Q_{L}|\,p_{1},\dots ,p_{n}\rangle ^{in}=\sum_{i=1}^{n}e^{L\theta
_{i}}|\,p_{1},\dots ,p_{n}\rangle ^{in}.  \label{J}
\end{equation}
In particular for $L=\pm 1$ the currents yield the energy momentum tensor $%
T^{\mu \nu }=T^{\nu \mu }$ with $\partial _{\mu }T^{\mu \nu }=0$.

The article is organized as follows: In section \ref{s2} we recall some
formulae of \cite{BFKZ,BK} and in particular those for breather form
factors, which we need in the following. The properties of the form factors
are translated to conditions for the `K-functions' and finally to simple
ones of the `p-functions'. In section \ref{s3} we investigate the asymptotic
behavior of bosonic form factors. In section \ref{s4} we discuss several
explicit examples of local operators as general exponentials of the
fundamental bose field, powers of the field, all higher conserved currents,
and the energy momentum tensor. Using induction and Liouville's theorem we
prove some identities which means that the same operators may be represented
in terms of different p-functions. These results are used in section \ref{s5}
to prove operator field equations as the quantum sine-Gordon field equation.
In section \ref{s6} we present further types of representations of
sine-Gordon breather form factors: a determinant formula (see also \cite
{Sm1,AlZa1,FMS,KM}) and two integral representation. One of them is new and
could presumably be generalized to other models with no backward scattering
(see also \cite{Fr}). A proof is delegated to the appendix.

\section{Breather form factors\label{s2}}

Using the bound state fusion method we derived in \cite{BK} from a general
formula for soliton anti-soliton form factors the pure breather form factor
formula which in particular for the case of lowest breathers may be written
in the form (\ref{1.1}) with $F(\theta )$ being the two particle form factor
function. It satisfies Watson's equations 
\[
F(\theta )=F(-\theta )S(\theta )=F(2\pi i-\theta ) 
\]
with the S-matrix given above. Explicitly it is given by the integral
representation \cite{KW} 
\begin{equation}
F(\theta )=\exp \int_{0}^{\infty }\frac{dt}{t}\,\frac{\cosh \left( \frac{1}{2%
}+\nu \right) t-\cosh \frac{1}{2}t}{\cosh \frac{1}{2}t\sinh t}\cosh t\left(
1-\frac{\theta }{i\pi }\right)  \label{2.1}
\end{equation}
normalized such that $F(\infty )=1$. In general form factors of one kind of
bosonic particles (i.e. with a diagonal S-matrix) satisfy the following
properties \cite{KW,Sm,BFKZ}.

\paragraph{Properties of the form factors:}

The form factor function $\mathcal{O}_{n}({\underline{\theta }})$ is
meromorphic with respect to all variables $\theta _{1},\dots ,\theta _{n}$.
It satisfies Watson's equations 
\begin{equation}
\mathcal{O}_{n}(\dots ,\theta _{i},\theta _{j},\dots )=\mathcal{O}_{n}(\dots
,\theta _{j},\theta _{i},\dots )\,S(\theta _{ij}).  \label{pf1}
\end{equation}
The crossing relation means for the connected part (see e.g. \cite{BK}) of
the matrix element 
\[
\langle \,p_{1}\,|\,\mathcal{O}(0)\,|\,p_{2},\dots ,p_{n}\,\rangle
_{conn.}^{in}=\,\mathcal{O}_{n}(\theta _{1}+i\pi ,\theta _{2},\dots ,\theta
_{n})=\mathcal{O}_{n}(\theta _{2},\dots ,\theta _{n},\theta _{1}-i\pi ) 
\]
which implies in particular 
\begin{equation}
\mathcal{O}_{n}(\theta _{1},\theta _{2},\dots ,\theta _{n},)=\mathcal{O}%
_{n}(\theta _{2},\dots ,\theta _{n},\theta _{1}-2\pi i)  \label{pf2}
\end{equation}
The function $\mathcal{O}_{n}({\underline{\theta }})$ has poles determined
by one-particle states in each sub-channel. In particular it has the
so-called annihilation poles at for example $\theta _{12}=i\pi $ such that
the recursion formula\footnote{%
This formula has been proposed in \cite{Sm} as a generalization of formulae
in \cite{KW} and it has been proven in \cite{BFKZ} using LSZ assumptions.}
is satisfied 
\begin{equation}
\mathop{\rm Res}_{\theta _{12}=i\pi }\mathcal{O}_{n}(\theta _{1},\dots
,\theta _{n})=2i\,\mathcal{O}_{n-2}(\theta _{3},\dots ,\theta _{n})\left( 
\mathbf{1}-S(\theta _{2n})\dots S(\theta _{23})\right) .  \label{pf3}
\end{equation}
Since we are dealing with relativistic quantum field theories Lorentz
covariance in the form 
\begin{equation}
\mathcal{O}_{n}(\theta _{1},\dots ,\theta _{n})=e^{-s\mu }\,\mathcal{O}%
_{n}(\theta _{1}+\mu ,\dots ,\theta _{n}+\mu )  \label{pf5}
\end{equation}
holds if the local operator transforms as $\mathcal{O}\to e^{s\mu }\mathcal{O%
}$ where $s$ is the ``spin'' of $\mathcal{O}$.

\paragraph{Conditions on the K-functions:}

Form factors of one kind of bosonic particles (which means that there is no
backward scattering) may be expressed by eq.~(\ref{1.1}) in terms of the
K-functions. Therefore properties of the form factors can be transformed to
the following relations: 
\begin{eqnarray}
&&K_{n}^{\mathcal{O}}(\dots ,\theta _{i},\theta _{j},\dots )=K_{n}^{\mathcal{%
O}}(\dots ,\theta _{j},\theta _{i},\dots )  \label{pk1} \\
&&K_{n}^{\mathcal{O}}(\underline{\theta })=K_{n}^{\mathcal{O}}(\theta
_{1}-2\pi i,\theta _{2},\dots ,\theta _{n})  \label{pk2} \\
&&\mathop{\rm Res}_{\theta _{12}=i\pi }K_{n}^{\mathcal{O}}(\underline{\theta 
})=\frac{2i}{F(i\pi )}\prod_{i=3}^{n}\frac{1}{F(\theta _{2i}+i\pi )F(\theta
_{2i})}\left( 1-\prod_{i=3}^{n}S(\theta _{2i})\right) K_{n-2}^{\mathcal{O}}(%
\underline{\theta }^{\prime \prime })  \label{pk3} \\
&&K_{n}^{\mathcal{O}}(\underline{\theta })=e^{-s\mu }\,K_{n}^{\mathcal{O}%
}(\theta _{1}+\mu ,\dots ,\theta _{n}+\mu )  \label{pk5}
\end{eqnarray}
where $\underline{\theta }=\theta _{1},\dots ,\theta _{n}$ and $\underline{%
\theta }^{\prime \prime }=\theta _{3},\dots ,\theta _{n}$.

\paragraph{Equations for the p-functions:}

Starting with a general integral representation for solitonic form factors
and using the bound state fusion method we have shown in \cite{BK} that the
lowest breather K-functions may be expressed by eq.~(\ref{1.2}) in terms of
breather p-functions which follow from solitonic p-functions. As already
mentioned in the introduction in the present article we make the \emph{%
Ansatz }that the K-function is of the form (\ref{1.2}) and we allow more
general breather p-functions. The \emph{Ansatz }(\ref{1.2}) transforms the
conditions on the K-function $K_{n}^{\mathcal{O}}(\underline{\theta })$ to
simpler equations for the p-function $p_{n}^{\mathcal{O}}(\underline{\theta }%
,\underline{l})$. The p-function $p_{n}^{\mathcal{O}}(\underline{\theta },%
\underline{l})$ is holomorphic with respect to all variables $\theta
_{1},\dots ,\theta _{n}$, is symmetric with respect to the exchange of the
variables $\theta _{i}$ and $l_{i}$ at the same time and is periodic with
period $2\pi i$. 
\begin{eqnarray}
p_{n}^{\mathcal{O}}(\dots ,\theta _{i},\theta _{j},\dots ,l_{i},l_{j},\dots
) &=&p_{n}^{\mathcal{O}}(\dots ,\theta _{j},\theta _{i},\dots
,l_{j},l_{i},\dots )  \label{pp1} \\
p_{n}^{\mathcal{O}}(\underline{\theta },\underline{l}) &=&p_{n}^{\mathcal{O}%
}(\theta _{1}-2\pi i,\theta _{2},\dots ,\theta _{n},\underline{l})
\label{pp2}
\end{eqnarray}
With the short hand notation $\underline{\theta }^{\prime }=\theta
_{2},\dots ,\theta _{n},\,\underline{\theta }^{\prime \prime }=\theta
_{3},\dots ,\theta _{n}$ and $\underline{l}^{\prime \prime }=l_{3},\dots
,l_{n}$ the recursion relation 
\begin{equation}
p_{n}^{\mathcal{O}}(\theta _{2}+i\pi ,\underline{\theta }^{\prime },%
\underline{l})=g(l_{1},l_{2})p_{n-2}^{\mathcal{O}}(\underline{\theta }%
^{\prime \prime },\underline{l}^{\prime \prime })+h(l_{1},l_{2})  \label{pp3}
\end{equation}
holds where $g(0,1)=g(1,0)=2/(F(i\pi )\sin \pi \nu )$ and $h(l_{1},l_{2})$
is independent of $\underline{l}^{\prime \prime }$. Lorentz covariance reads
as 
\begin{equation}
p_{n}^{\mathcal{O}}(\theta _{1}+\mu ,\dots ,\theta _{n}+\mu ,\underline{l}%
)=e^{s\mu }\,p_{n}^{\mathcal{O}}(\theta _{1},\dots ,\theta _{n},\underline{l}%
).  \label{pp5}
\end{equation}
We now show that these conditions of the p-function are sufficient to
guarantee the properties of the form factors.

\begin{theorem}
\label{t1}If the p-function $p_{n}^{\mathcal{O}}(\underline{\theta },%
\underline{l})$ satisfies the conditions (\ref{pp1}--\ref{pp5}), the
K-function $K_{n}^{\mathcal{O}}(\underline{\theta })$ satisfies the
conditions (\ref{pk1}--\ref{pk5}) and therefore the form factor function $%
\mathcal{O}_{n}(\underline{\theta })$ satisfies the properties (\ref{pf1}--%
\ref{pf5}).
\end{theorem}

\proof   %
Except for (\ref{pk3}) all claims are obvious. Taking the residue of (\ref
{1.2}) and inserting (\ref{pk3}) we obtain (with $a=i\sin \pi \nu )$%
\begin{multline*}
\mathop{\rm Res}_{\theta _{12}=i\pi }K_{n}(\underline{\theta }%
)=-a\sum_{l_{3}=0}^{1}\dots \sum_{l_{r}=0}^{1}(-1)^{l_{3}+\dots
+l_{n}}\prod_{3=i<j}^{n}\left( 1+\frac{l_{i}-l_{j}}{\sinh \theta _{ij}}%
a\right) \\
\times \sum_{l_{1}\neq
l_{2}}(-1)^{l_{1}+l_{2}}(l_{1}-l_{2})\prod_{i=3}^{n}\left( \left( 1+\frac{%
l_{1}-l_{i}}{\sinh (\theta _{2i}+i\pi )}a\right) \left( 1+\frac{l_{2}-l_{i}}{%
\sinh \theta _{2i}}a\right) \right) \\
\times \left( \,g(l_{1},l_{2})p_{n-2}^{\mathcal{O}}(\underline{\theta }%
^{\prime \prime },\underline{l}^{\prime \prime })+h(l_{1},l_{2})\right) \\
=\frac{2i}{F(i\pi )}K_{n-2}(\underline{\theta }^{\prime \prime })\left(
\prod_{i=3}^{n}\left( 1+\frac{a}{\sinh \theta _{2i}}\right)
-\prod_{i=3}^{n}\left( 1-\frac{a}{\sinh \theta _{2i}}\right) \right) +h\text{%
-term.}
\end{multline*}
We have used the identity 
\begin{multline*}
\sum_{l_{1}\neq l_{2}}(-1)^{l_{1}+l_{2}}(l_{1}-l_{2})\prod_{i=3}^{n}\left(
\left( 1+\frac{l_{1}-l_{i}}{\sinh (\theta _{2i}+i\pi )}a\right) \left( 1+%
\frac{l_{2}-l_{i}}{\sinh \theta _{2i}}a\right) \right) g(l_{1},l_{2}) \\
=g(0,1)\prod_{i=3}^{n}\left( 1+\frac{a}{\sinh \theta _{2i}}\right)
-g(1,0)\prod_{i=3}^{n}\left( 1-\frac{a}{\sinh \theta _{2i}}\right)
\end{multline*}
valid for all $l_{i}\,(i\geq 3)$. The same relation is valid when we replace 
$g$ by $h$. The equation (\ref{pk3}) now follows from $g(0,1)=g(1,0)=2/(F(i%
\pi )\sin \pi \nu )$ and 
\[
F(\theta +i\pi )F(\theta )=1/\left( 1-\frac{i\sin \pi \nu }{\sinh \theta }%
\right) 
\]
(which is easily obtained from the integral representation (\ref{2.1})),
provided that the $h$-term vanishes. This is a consequence of the following
lemma. Note that the $h$-term is proportional to a $K_{n-2}(\underline{%
\theta }^{\prime \prime })$ given by the formula (\ref{1.2}) with a
p-function independent of the $\underline{l}^{\prime \prime }$.%
\endproof   %

\begin{lemma}
\label{l1}If the 'p-function' in (\ref{1.2}) does not depend on $l_{1},\dots
,l_{n}$ then the corresponding K-function vanishes.
\end{lemma}

\proof   %
The proof is easy and obtained by using induction and Liouville's theorem:
We easily obtain $K_{1}(\underline{\theta })=K_{2}(\underline{\theta })=0$.
As induction assumptions we take $K_{n-2}(\underline{\theta }^{\prime \prime
})=0$. The function $K_{n}(\underline{\theta })$ is a meromorphic functions
in terms of the $x_{i}=e^{\theta _{i}}$ with at most simple poles at $%
x_{i}=\pm x_{j}$ since $\sinh \theta
_{ij}=(x_{i}+x_{j})(x_{i}-x_{j})/(2x_{i}x_{j})$. The residues of the poles
at $x_{i}=x_{j}$ vanish because of the symmetry. Furthermore the residues at 
$x_{i}=-x_{j}$ are proportional to $K_{n-2}(\underline{\theta }^{\prime
\prime })$ because similarly to the proof of theorem \ref{t1} we have 
\[
\mathop{\rm Res}_{\theta _{12}=i\pi }K_{n}(\underline{\theta })=aK_{n-2}(%
\underline{\theta }^{\prime \prime })\left( \prod_{i=3}^{n}\left( 1+\frac{a}{%
\sinh \theta _{2i}}\right) -\prod_{i=3}^{n}\left( 1-\frac{a}{\sinh \theta
_{2i}}\right) \right) 
\]
Therefore the function $K_{n}(\underline{\theta })$ is holomorphic
everywhere. Furthermore for $x_{1}\rightarrow \infty $ we have the
asymptotic behavior 
\begin{multline}
K_{n}(\underline{\theta })=\sum_{l_{2}=0}^{1}\dots
\sum_{l_{n}=0}^{1}(-1)^{l_{2}+\dots +l_{n}}\prod_{2\leq i<j\leq n}\left(
1+(l_{i}-l_{j})\frac{i\sin \pi \nu }{\sinh \theta _{ij}}\right) \\
\times \sum_{l_{1}=0}^{1}(-1)^{l_{1}}\prod_{j=2}^{n}\left( 1+(l_{1}-l_{j})%
\frac{i\sin \pi \nu }{\sinh \theta _{1j}}\right) \rightarrow 0  \label{as}
\end{multline}
Therefore $K_{n}(\underline{\theta })$ vanishes identically by Liouville's
theorem.%
\endproof   %

\section{Asymptotic behavior of bosonic form factors\label{s3}}

In this section we derive the asymptotic behavior of bosonic form factors by
mean of general techniques of renormalized local quantum field theory. In
particular we use perturbation theory in term of Feynman graphs. As the
simplest example we investigate first the asymptotic behavior for $%
p_{1}\rightarrow \infty $ or $\theta _{12}\rightarrow \infty $ of 
\begin{eqnarray*}
\left\langle \,0\,\,|\!:\!\varphi ^{2}\!:\!\,|\,p_{1},p_{2}\,\right\rangle
^{in} &=&2\left\langle \,0\,|\,\varphi (0)\,|\,p_{1}\right\rangle
\left\langle \,0\,\left| \,\varphi (0)\,\right| \,p_{2}\,\right\rangle +o(1)
\\
&=&2Z^{\varphi }+o(1)
\end{eqnarray*}
where $:\dots :$ means normal ordering with respect to the physical vacuum.
This may be seen in perturbation theory as follows: Feynman graph expansion
in lowest order means 
\begin{align*}
\langle \,0\,|\!& :\!\varphi ^{2}\!:\!\,|\,p_{1},p_{2}\,\rangle
^{in}=2\left( 
\begin{array}{c}
\unitlength4mm\begin{picture}(8,4) \put(1,3){\makebox(0,0){$\bullet$}}
\put(6,3){\makebox(0,0){$\bullet$}} \put(1,3.8){\makebox(0,0){$\varphi^2$}}
\put(6,3.8){\makebox(0,0){$\varphi^2$}} \put(0,0){\line(1,3){1}}
\put(2,0){\line(-1,3){1}} \put(3,1.5){$+$} \put(5,0){\line(1,1){1}}
\put(7,0){\line(-1,1){1}} \put(6,2){\oval(2,2)} \end{picture}
\end{array}
\right) +O(\beta ^{4}) \\
& =2\left( 1+i\alpha \beta ^{2}\tfrac{1}{2}\int \frac{d^{2}k}{(2\pi )^{2}}%
\frac{i}{k^{2}-m_{1}^{2}}\frac{i}{(p-k)^{2}-m_{1}^{2}}\right) +O(\beta ^{4})
\\
& =2+\frac{\beta ^{2}}{4\pi }\frac{i\pi -\theta _{12}}{\sinh \theta _{12}}%
+O(\beta ^{4})
\end{align*}
The second graph is of order $O(\ln p_{1}/p_{1})$ for $p_{1}\to \infty $.
This is typical also for all orders in perturbation theory:

\[
\begin{array}{c}
\unitlength4mm\begin{picture}(4,6) \put(2,2){\oval(4,2)}
\put(2,5){\makebox(0,0){$\bullet$}}
\put(2,5.8){\makebox(0,0){$:\varphi^2:$}} \put(1,3){\line(1,2){1}}
\put(3,3){\line(-1,2){1}} \put(1,0){\line(0,1){1}} \put(3,0){\line(0,1){1}}
\put(-.5,0){$\theta_1$} \put(3.4,0){$\theta_2$} \end{picture}
\end{array}
~=~ 
\begin{array}{c}
\unitlength4mm\begin{picture}(6,6) \put(1,2){\oval(2,2)}
\put(5,2){\oval(2,2)} \put(3,5){\makebox(0,0){$\bullet$}}
\put(3,5.8){\makebox(0,0){$:\varphi^2:$}} \put(1,3){\line(1,1){2}}
\put(5,3){\line(-1,1){2}} \put(1,0){\line(0,1){1}} \put(5,0){\line(0,1){1}}
\put(-.1,0){$\theta_1$} \put(5.4,0){$\theta_2$} \end{picture}
\end{array}
~+~ 
\begin{array}{c}
\unitlength4mm\begin{picture}(6,6) \put(1,2){\oval(2,2)}
\put(5,2){\oval(2,2)} \put(3,5){\makebox(0,0){$\bullet$}}
\put(3,5.8){\makebox(0,0){$:\varphi^2:$}} \put(1,3){\line(1,1){2}}
\put(5,3){\line(-1,1){2}} \put(1,0){\line(0,1){1}} \put(5,0){\line(0,1){1}}
\put(-.1,0){$\theta_1$} \put(5.4,0){$\theta_2$}
\put(1.9,1.5){\line(1,0){2.2}} \put(1.9,2.5){\line(1,0){2.2}} \end{picture}
\end{array}
~+\cdots 
\]
Weinberg's power counting theorem says that the second term and also all
higher terms where more lines connect the two bubbles are again at least of
order $O(\ln p_{1}/p_{1})$ for $p_{1}\to \infty $.

The wave function renormalization constant $Z^{\varphi }$ is defined as
usual by the two-point function of the (unrenormalized) field 
\begin{align*}
\int \langle \,0\left| T\varphi (x)\varphi (0)\right| 0\,\rangle
\,e^{ipx}\,d^{2}x& = 
\begin{array}{l}
\unitlength4mm\begin{picture}(16,3) \put(0,1){\line(1,0){2}}
\put(2.8,.8){$+$} \put(4,1){\line(1,0){3}} \put(5.5,1){\oval(1,1)}
\put(7.8,.8){$+$} \put(9,1){\line(1,0){5}} \put(10.5,1){\oval(1,1)}
\put(12.5,1){\oval(1,1)} \put(15,.8){$+\cdots$} \end{picture}
\end{array}
\\
& =\dfrac{i}{p^{2}-\alpha -\Pi (p^{2})}=\dfrac{iZ^{\varphi }}{%
p^{2}-m^{2}-\Pi _{ren}(p^{2})}
\end{align*}
where $\Pi (p)$ is the self energy, which means that it is given by the sum
of all amputated one-particle-irreducible graphs 
\[
-i\Pi (p^{2})= 
\begin{array}{c}
\unitlength4mm \begin{picture}(3,2)
\put(0,1){\line(1,0){.5}}\put(2.5,1){\line(1,0){.5}}
\put(.5,0){\framebox(2,2){}} \end{picture}
\end{array}
\]
The physical breather mass $m$, the wave function renormalization constant $%
Z^{\varphi }$ and the renormalized breather self energy are given by 
\begin{eqnarray*}
m^{2} &=&\alpha +\Pi (m^{2}) \\
\frac{1}{Z^{\varphi }} &=&1-\Pi ^{\prime }(m^{2}) \\
\Pi _{ren}(p^{2}) &=&Z^{\varphi }\left( \Pi (p^{2})-\Pi (m^{2})-\left(
p^{2}-m^{2}\right) \Pi ^{\prime }(m^{2})\right) .
\end{eqnarray*}
Since the sine-Gordon model is a 'super renormalizable quantum field theory'
both renormalization constants $\Pi (m^{2})$ and $\Pi ^{\prime }(m^{2})$
become finite after taking normal ordering in the interaction Lagrangian..
They can be calculated exactly. The wave function renormalization constant
was obtained in \cite{KW} 
\begin{equation}
Z^{\varphi }=(1+\nu )\frac{\frac{\pi }{2}\nu }{\sin \frac{\pi }{2}\nu }\exp
\left( -\frac{1}{\pi }\int_{0}^{\pi \nu }\frac{t}{\sin t}dt\right)  \label{Z}
\end{equation}
and the relation of the unrenormalized and the physical mass is calculated
in the present article (see section \ref{s5}) 
\[
\alpha =m^{2}\frac{\pi \nu }{\sin \pi \nu }\,. 
\]
Both relations have been checked in perturbation theory.

\begin{figure}[tbh]
\[
\setlength{\unitlength}{0.240900pt} \ifx\plotpoint\undefined%
\newsavebox{\plotpoint}\fi
\begin{picture}(740,450)(0,0)
\font\gnuplot=cmr10 at 10pt
\gnuplot
\sbox{\plotpoint}{\rule[-0.200pt]{0.400pt}{0.400pt}}%
\put(160.0,123.0){\rule[-0.200pt]{4.818pt}{0.400pt}}
\put(140,123){\makebox(0,0)[r]{0.994}}
\put(669.0,123.0){\rule[-0.200pt]{4.818pt}{0.400pt}}
\put(160.0,171.0){\rule[-0.200pt]{4.818pt}{0.400pt}}
\put(669.0,171.0){\rule[-0.200pt]{4.818pt}{0.400pt}}
\put(160.0,219.0){\rule[-0.200pt]{4.818pt}{0.400pt}}
\put(140,219){\makebox(0,0)[r]{0.996}}
\put(669.0,219.0){\rule[-0.200pt]{4.818pt}{0.400pt}}
\put(160.0,267.0){\rule[-0.200pt]{4.818pt}{0.400pt}}
\put(669.0,267.0){\rule[-0.200pt]{4.818pt}{0.400pt}}
\put(160.0,314.0){\rule[-0.200pt]{4.818pt}{0.400pt}}
\put(140,314){\makebox(0,0)[r]{0.998}}
\put(669.0,314.0){\rule[-0.200pt]{4.818pt}{0.400pt}}
\put(160.0,362.0){\rule[-0.200pt]{4.818pt}{0.400pt}}
\put(669.0,362.0){\rule[-0.200pt]{4.818pt}{0.400pt}}
\put(160.0,410.0){\rule[-0.200pt]{4.818pt}{0.400pt}}
\put(140,410){\makebox(0,0)[r]{1}}
\put(669.0,410.0){\rule[-0.200pt]{4.818pt}{0.400pt}}
\put(160.0,123.0){\rule[-0.200pt]{0.400pt}{4.818pt}}
\put(160,82){\makebox(0,0){-1}}
\put(160.0,390.0){\rule[-0.200pt]{0.400pt}{4.818pt}}
\put(248.0,123.0){\rule[-0.200pt]{0.400pt}{4.818pt}}
\put(248,82){\makebox(0,0){-0.8}}
\put(248.0,390.0){\rule[-0.200pt]{0.400pt}{4.818pt}}
\put(336.0,123.0){\rule[-0.200pt]{0.400pt}{4.818pt}}
\put(336,82){\makebox(0,0){-0.6}}
\put(336.0,390.0){\rule[-0.200pt]{0.400pt}{4.818pt}}
\put(424.0,123.0){\rule[-0.200pt]{0.400pt}{4.818pt}}
\put(424,82){\makebox(0,0){-0.4}}
\put(424.0,390.0){\rule[-0.200pt]{0.400pt}{4.818pt}}
\put(513.0,123.0){\rule[-0.200pt]{0.400pt}{4.818pt}}
\put(513,82){\makebox(0,0){-0.2}}
\put(513.0,390.0){\rule[-0.200pt]{0.400pt}{4.818pt}}
\put(601.0,123.0){\rule[-0.200pt]{0.400pt}{4.818pt}}
\put(601,82){\makebox(0,0){0}}
\put(601.0,390.0){\rule[-0.200pt]{0.400pt}{4.818pt}}
\put(689.0,123.0){\rule[-0.200pt]{0.400pt}{4.818pt}}
\put(689,82){\makebox(0,0){0.2}}
\put(689.0,390.0){\rule[-0.200pt]{0.400pt}{4.818pt}}
\put(160.0,123.0){\rule[-0.200pt]{127.436pt}{0.400pt}}
\put(689.0,123.0){\rule[-0.200pt]{0.400pt}{69.138pt}}
\put(160.0,410.0){\rule[-0.200pt]{127.436pt}{0.400pt}}
\put(424,0){\makebox(0,0){(a)}}
\put(160.0,123.0){\rule[-0.200pt]{0.400pt}{69.138pt}}
\put(160,410){\usebox{\plotpoint}}
\put(160,408.17){\rule{1.900pt}{0.400pt}}
\multiput(160.00,409.17)(5.056,-2.000){2}{\rule{0.950pt}{0.400pt}}
\multiput(169.00,406.94)(1.212,-0.468){5}{\rule{1.000pt}{0.113pt}}
\multiput(169.00,407.17)(6.924,-4.000){2}{\rule{0.500pt}{0.400pt}}
\multiput(178.00,402.93)(0.569,-0.485){11}{\rule{0.557pt}{0.117pt}}
\multiput(178.00,403.17)(6.844,-7.000){2}{\rule{0.279pt}{0.400pt}}
\multiput(186.00,395.93)(0.495,-0.489){15}{\rule{0.500pt}{0.118pt}}
\multiput(186.00,396.17)(7.962,-9.000){2}{\rule{0.250pt}{0.400pt}}
\multiput(195.59,385.56)(0.489,-0.611){15}{\rule{0.118pt}{0.589pt}}
\multiput(194.17,386.78)(9.000,-9.778){2}{\rule{0.400pt}{0.294pt}}
\multiput(204.59,374.37)(0.489,-0.669){15}{\rule{0.118pt}{0.633pt}}
\multiput(203.17,375.69)(9.000,-10.685){2}{\rule{0.400pt}{0.317pt}}
\multiput(213.59,362.00)(0.489,-0.786){15}{\rule{0.118pt}{0.722pt}}
\multiput(212.17,363.50)(9.000,-12.501){2}{\rule{0.400pt}{0.361pt}}
\multiput(222.59,348.00)(0.489,-0.786){15}{\rule{0.118pt}{0.722pt}}
\multiput(221.17,349.50)(9.000,-12.501){2}{\rule{0.400pt}{0.361pt}}
\multiput(231.59,333.68)(0.488,-0.890){13}{\rule{0.117pt}{0.800pt}}
\multiput(230.17,335.34)(8.000,-12.340){2}{\rule{0.400pt}{0.400pt}}
\multiput(239.59,319.82)(0.489,-0.844){15}{\rule{0.118pt}{0.767pt}}
\multiput(238.17,321.41)(9.000,-13.409){2}{\rule{0.400pt}{0.383pt}}
\multiput(248.59,305.00)(0.489,-0.786){15}{\rule{0.118pt}{0.722pt}}
\multiput(247.17,306.50)(9.000,-12.501){2}{\rule{0.400pt}{0.361pt}}
\multiput(257.59,290.82)(0.489,-0.844){15}{\rule{0.118pt}{0.767pt}}
\multiput(256.17,292.41)(9.000,-13.409){2}{\rule{0.400pt}{0.383pt}}
\multiput(266.59,276.00)(0.489,-0.786){15}{\rule{0.118pt}{0.722pt}}
\multiput(265.17,277.50)(9.000,-12.501){2}{\rule{0.400pt}{0.361pt}}
\multiput(275.59,261.68)(0.488,-0.890){13}{\rule{0.117pt}{0.800pt}}
\multiput(274.17,263.34)(8.000,-12.340){2}{\rule{0.400pt}{0.400pt}}
\multiput(283.59,248.19)(0.489,-0.728){15}{\rule{0.118pt}{0.678pt}}
\multiput(282.17,249.59)(9.000,-11.593){2}{\rule{0.400pt}{0.339pt}}
\multiput(292.59,235.37)(0.489,-0.669){15}{\rule{0.118pt}{0.633pt}}
\multiput(291.17,236.69)(9.000,-10.685){2}{\rule{0.400pt}{0.317pt}}
\multiput(301.59,223.56)(0.489,-0.611){15}{\rule{0.118pt}{0.589pt}}
\multiput(300.17,224.78)(9.000,-9.778){2}{\rule{0.400pt}{0.294pt}}
\multiput(310.59,212.74)(0.489,-0.553){15}{\rule{0.118pt}{0.544pt}}
\multiput(309.17,213.87)(9.000,-8.870){2}{\rule{0.400pt}{0.272pt}}
\multiput(319.00,203.93)(0.495,-0.489){15}{\rule{0.500pt}{0.118pt}}
\multiput(319.00,204.17)(7.962,-9.000){2}{\rule{0.250pt}{0.400pt}}
\multiput(328.00,194.93)(0.569,-0.485){11}{\rule{0.557pt}{0.117pt}}
\multiput(328.00,195.17)(6.844,-7.000){2}{\rule{0.279pt}{0.400pt}}
\multiput(336.00,187.93)(0.645,-0.485){11}{\rule{0.614pt}{0.117pt}}
\multiput(336.00,188.17)(7.725,-7.000){2}{\rule{0.307pt}{0.400pt}}
\multiput(345.00,180.93)(0.933,-0.477){7}{\rule{0.820pt}{0.115pt}}
\multiput(345.00,181.17)(7.298,-5.000){2}{\rule{0.410pt}{0.400pt}}
\multiput(354.00,175.95)(1.802,-0.447){3}{\rule{1.300pt}{0.108pt}}
\multiput(354.00,176.17)(6.302,-3.000){2}{\rule{0.650pt}{0.400pt}}
\multiput(363.00,172.95)(1.802,-0.447){3}{\rule{1.300pt}{0.108pt}}
\multiput(363.00,173.17)(6.302,-3.000){2}{\rule{0.650pt}{0.400pt}}
\multiput(389.00,171.61)(1.802,0.447){3}{\rule{1.300pt}{0.108pt}}
\multiput(389.00,170.17)(6.302,3.000){2}{\rule{0.650pt}{0.400pt}}
\multiput(398.00,174.61)(1.802,0.447){3}{\rule{1.300pt}{0.108pt}}
\multiput(398.00,173.17)(6.302,3.000){2}{\rule{0.650pt}{0.400pt}}
\multiput(407.00,177.59)(0.933,0.477){7}{\rule{0.820pt}{0.115pt}}
\multiput(407.00,176.17)(7.298,5.000){2}{\rule{0.410pt}{0.400pt}}
\multiput(416.00,182.59)(0.645,0.485){11}{\rule{0.614pt}{0.117pt}}
\multiput(416.00,181.17)(7.725,7.000){2}{\rule{0.307pt}{0.400pt}}
\multiput(425.00,189.59)(0.569,0.485){11}{\rule{0.557pt}{0.117pt}}
\multiput(425.00,188.17)(6.844,7.000){2}{\rule{0.279pt}{0.400pt}}
\multiput(433.00,196.59)(0.495,0.489){15}{\rule{0.500pt}{0.118pt}}
\multiput(433.00,195.17)(7.962,9.000){2}{\rule{0.250pt}{0.400pt}}
\multiput(442.59,205.00)(0.489,0.553){15}{\rule{0.118pt}{0.544pt}}
\multiput(441.17,205.00)(9.000,8.870){2}{\rule{0.400pt}{0.272pt}}
\multiput(451.59,215.00)(0.489,0.611){15}{\rule{0.118pt}{0.589pt}}
\multiput(450.17,215.00)(9.000,9.778){2}{\rule{0.400pt}{0.294pt}}
\multiput(460.59,226.00)(0.489,0.669){15}{\rule{0.118pt}{0.633pt}}
\multiput(459.17,226.00)(9.000,10.685){2}{\rule{0.400pt}{0.317pt}}
\multiput(469.59,238.00)(0.488,0.824){13}{\rule{0.117pt}{0.750pt}}
\multiput(468.17,238.00)(8.000,11.443){2}{\rule{0.400pt}{0.375pt}}
\multiput(477.59,251.00)(0.489,0.786){15}{\rule{0.118pt}{0.722pt}}
\multiput(476.17,251.00)(9.000,12.501){2}{\rule{0.400pt}{0.361pt}}
\multiput(486.59,265.00)(0.489,0.786){15}{\rule{0.118pt}{0.722pt}}
\multiput(485.17,265.00)(9.000,12.501){2}{\rule{0.400pt}{0.361pt}}
\multiput(495.59,279.00)(0.489,0.844){15}{\rule{0.118pt}{0.767pt}}
\multiput(494.17,279.00)(9.000,13.409){2}{\rule{0.400pt}{0.383pt}}
\multiput(504.59,294.00)(0.489,0.786){15}{\rule{0.118pt}{0.722pt}}
\multiput(503.17,294.00)(9.000,12.501){2}{\rule{0.400pt}{0.361pt}}
\multiput(513.59,308.00)(0.488,0.956){13}{\rule{0.117pt}{0.850pt}}
\multiput(512.17,308.00)(8.000,13.236){2}{\rule{0.400pt}{0.425pt}}
\multiput(521.59,323.00)(0.489,0.786){15}{\rule{0.118pt}{0.722pt}}
\multiput(520.17,323.00)(9.000,12.501){2}{\rule{0.400pt}{0.361pt}}
\multiput(530.59,337.00)(0.489,0.786){15}{\rule{0.118pt}{0.722pt}}
\multiput(529.17,337.00)(9.000,12.501){2}{\rule{0.400pt}{0.361pt}}
\multiput(539.59,351.00)(0.489,0.786){15}{\rule{0.118pt}{0.722pt}}
\multiput(538.17,351.00)(9.000,12.501){2}{\rule{0.400pt}{0.361pt}}
\multiput(548.59,365.00)(0.489,0.669){15}{\rule{0.118pt}{0.633pt}}
\multiput(547.17,365.00)(9.000,10.685){2}{\rule{0.400pt}{0.317pt}}
\multiput(557.59,377.00)(0.489,0.611){15}{\rule{0.118pt}{0.589pt}}
\multiput(556.17,377.00)(9.000,9.778){2}{\rule{0.400pt}{0.294pt}}
\multiput(566.59,388.00)(0.488,0.560){13}{\rule{0.117pt}{0.550pt}}
\multiput(565.17,388.00)(8.000,7.858){2}{\rule{0.400pt}{0.275pt}}
\multiput(574.00,397.59)(0.645,0.485){11}{\rule{0.614pt}{0.117pt}}
\multiput(574.00,396.17)(7.725,7.000){2}{\rule{0.307pt}{0.400pt}}
\multiput(583.00,404.60)(1.212,0.468){5}{\rule{1.000pt}{0.113pt}}
\multiput(583.00,403.17)(6.924,4.000){2}{\rule{0.500pt}{0.400pt}}
\put(592,408.17){\rule{1.900pt}{0.400pt}}
\multiput(592.00,407.17)(5.056,2.000){2}{\rule{0.950pt}{0.400pt}}
\put(601,408.17){\rule{1.900pt}{0.400pt}}
\multiput(601.00,409.17)(5.056,-2.000){2}{\rule{0.950pt}{0.400pt}}
\multiput(610.00,406.93)(0.821,-0.477){7}{\rule{0.740pt}{0.115pt}}
\multiput(610.00,407.17)(6.464,-5.000){2}{\rule{0.370pt}{0.400pt}}
\multiput(618.59,400.56)(0.489,-0.611){15}{\rule{0.118pt}{0.589pt}}
\multiput(617.17,401.78)(9.000,-9.778){2}{\rule{0.400pt}{0.294pt}}
\multiput(627.59,388.82)(0.489,-0.844){15}{\rule{0.118pt}{0.767pt}}
\multiput(626.17,390.41)(9.000,-13.409){2}{\rule{0.400pt}{0.383pt}}
\multiput(636.59,372.71)(0.489,-1.194){15}{\rule{0.118pt}{1.033pt}}
\multiput(635.17,374.86)(9.000,-18.855){2}{\rule{0.400pt}{0.517pt}}
\multiput(645.59,350.60)(0.489,-1.543){15}{\rule{0.118pt}{1.300pt}}
\multiput(644.17,353.30)(9.000,-24.302){2}{\rule{0.400pt}{0.650pt}}
\multiput(654.59,322.13)(0.489,-2.009){15}{\rule{0.118pt}{1.656pt}}
\multiput(653.17,325.56)(9.000,-31.564){2}{\rule{0.400pt}{0.828pt}}
\multiput(663.59,284.87)(0.488,-2.739){13}{\rule{0.117pt}{2.200pt}}
\multiput(662.17,289.43)(8.000,-37.434){2}{\rule{0.400pt}{1.100pt}}
\multiput(671.59,241.99)(0.489,-2.999){15}{\rule{0.118pt}{2.411pt}}
\multiput(670.17,247.00)(9.000,-46.996){2}{\rule{0.400pt}{1.206pt}}
\multiput(680.59,188.33)(0.489,-3.524){15}{\rule{0.118pt}{2.811pt}}
\multiput(679.17,194.17)(9.000,-55.165){2}{\rule{0.400pt}{1.406pt}}
\put(372.0,171.0){\rule[-0.200pt]{4.095pt}{0.400pt}}
\put(689,139){\usebox{\plotpoint}}
\end{picture}
\setlength{\unitlength}{0.240900pt} \ifx\plotpoint\undefined%
\newsavebox{\plotpoint}\fi
\begin{picture}(629,450)(0,0)
\font\gnuplot=cmr10 at 10pt
\gnuplot
\sbox{\plotpoint}{\rule[-0.200pt]{0.400pt}{0.400pt}}%
\put(120.0,123.0){\rule[-0.200pt]{4.818pt}{0.400pt}}
\put(100,123){\makebox(0,0)[r]{0}}
\put(588.0,123.0){\rule[-0.200pt]{4.818pt}{0.400pt}}
\put(120.0,180.0){\rule[-0.200pt]{4.818pt}{0.400pt}}
\put(100,180){\makebox(0,0)[r]{0.2}}
\put(588.0,180.0){\rule[-0.200pt]{4.818pt}{0.400pt}}
\put(120.0,238.0){\rule[-0.200pt]{4.818pt}{0.400pt}}
\put(100,238){\makebox(0,0)[r]{0.4}}
\put(588.0,238.0){\rule[-0.200pt]{4.818pt}{0.400pt}}
\put(120.0,295.0){\rule[-0.200pt]{4.818pt}{0.400pt}}
\put(100,295){\makebox(0,0)[r]{0.6}}
\put(588.0,295.0){\rule[-0.200pt]{4.818pt}{0.400pt}}
\put(120.0,353.0){\rule[-0.200pt]{4.818pt}{0.400pt}}
\put(100,353){\makebox(0,0)[r]{0.8}}
\put(588.0,353.0){\rule[-0.200pt]{4.818pt}{0.400pt}}
\put(120.0,410.0){\rule[-0.200pt]{4.818pt}{0.400pt}}
\put(100,410){\makebox(0,0)[r]{1}}
\put(588.0,410.0){\rule[-0.200pt]{4.818pt}{0.400pt}}
\put(120.0,123.0){\rule[-0.200pt]{0.400pt}{4.818pt}}
\put(120,82){\makebox(0,0){0}}
\put(120.0,390.0){\rule[-0.200pt]{0.400pt}{4.818pt}}
\put(218.0,123.0){\rule[-0.200pt]{0.400pt}{4.818pt}}
\put(218,82){\makebox(0,0){0.2}}
\put(218.0,390.0){\rule[-0.200pt]{0.400pt}{4.818pt}}
\put(315.0,123.0){\rule[-0.200pt]{0.400pt}{4.818pt}}
\put(315,82){\makebox(0,0){0.4}}
\put(315.0,390.0){\rule[-0.200pt]{0.400pt}{4.818pt}}
\put(413.0,123.0){\rule[-0.200pt]{0.400pt}{4.818pt}}
\put(413,82){\makebox(0,0){0.6}}
\put(413.0,390.0){\rule[-0.200pt]{0.400pt}{4.818pt}}
\put(510.0,123.0){\rule[-0.200pt]{0.400pt}{4.818pt}}
\put(510,82){\makebox(0,0){0.8}}
\put(510.0,390.0){\rule[-0.200pt]{0.400pt}{4.818pt}}
\put(608.0,123.0){\rule[-0.200pt]{0.400pt}{4.818pt}}
\put(608,82){\makebox(0,0){1}}
\put(608.0,390.0){\rule[-0.200pt]{0.400pt}{4.818pt}}
\put(120.0,123.0){\rule[-0.200pt]{117.559pt}{0.400pt}}
\put(608.0,123.0){\rule[-0.200pt]{0.400pt}{69.138pt}}
\put(120.0,410.0){\rule[-0.200pt]{117.559pt}{0.400pt}}
\put(364,0){\makebox(0,0){(b)}}
\put(120.0,123.0){\rule[-0.200pt]{0.400pt}{69.138pt}}
\put(120,410){\usebox{\plotpoint}}
\put(179,408.67){\rule{2.168pt}{0.400pt}}
\multiput(179.00,409.17)(4.500,-1.000){2}{\rule{1.084pt}{0.400pt}}
\put(120.0,410.0){\rule[-0.200pt]{14.213pt}{0.400pt}}
\put(208,407.67){\rule{2.409pt}{0.400pt}}
\multiput(208.00,408.17)(5.000,-1.000){2}{\rule{1.204pt}{0.400pt}}
\put(188.0,409.0){\rule[-0.200pt]{4.818pt}{0.400pt}}
\put(227,406.67){\rule{2.409pt}{0.400pt}}
\multiput(227.00,407.17)(5.000,-1.000){2}{\rule{1.204pt}{0.400pt}}
\put(218.0,408.0){\rule[-0.200pt]{2.168pt}{0.400pt}}
\put(247,405.67){\rule{2.409pt}{0.400pt}}
\multiput(247.00,406.17)(5.000,-1.000){2}{\rule{1.204pt}{0.400pt}}
\put(257,404.67){\rule{2.168pt}{0.400pt}}
\multiput(257.00,405.17)(4.500,-1.000){2}{\rule{1.084pt}{0.400pt}}
\put(237.0,407.0){\rule[-0.200pt]{2.409pt}{0.400pt}}
\put(276,403.67){\rule{2.409pt}{0.400pt}}
\multiput(276.00,404.17)(5.000,-1.000){2}{\rule{1.204pt}{0.400pt}}
\put(286,402.17){\rule{2.100pt}{0.400pt}}
\multiput(286.00,403.17)(5.641,-2.000){2}{\rule{1.050pt}{0.400pt}}
\put(296,400.67){\rule{2.168pt}{0.400pt}}
\multiput(296.00,401.17)(4.500,-1.000){2}{\rule{1.084pt}{0.400pt}}
\put(305,399.67){\rule{2.409pt}{0.400pt}}
\multiput(305.00,400.17)(5.000,-1.000){2}{\rule{1.204pt}{0.400pt}}
\put(315,398.17){\rule{2.100pt}{0.400pt}}
\multiput(315.00,399.17)(5.641,-2.000){2}{\rule{1.050pt}{0.400pt}}
\put(325,396.17){\rule{2.100pt}{0.400pt}}
\multiput(325.00,397.17)(5.641,-2.000){2}{\rule{1.050pt}{0.400pt}}
\put(335,394.17){\rule{1.900pt}{0.400pt}}
\multiput(335.00,395.17)(5.056,-2.000){2}{\rule{0.950pt}{0.400pt}}
\put(344,392.17){\rule{2.100pt}{0.400pt}}
\multiput(344.00,393.17)(5.641,-2.000){2}{\rule{1.050pt}{0.400pt}}
\put(354,390.17){\rule{2.100pt}{0.400pt}}
\multiput(354.00,391.17)(5.641,-2.000){2}{\rule{1.050pt}{0.400pt}}
\multiput(364.00,388.95)(2.025,-0.447){3}{\rule{1.433pt}{0.108pt}}
\multiput(364.00,389.17)(7.025,-3.000){2}{\rule{0.717pt}{0.400pt}}
\multiput(374.00,385.95)(2.025,-0.447){3}{\rule{1.433pt}{0.108pt}}
\multiput(374.00,386.17)(7.025,-3.000){2}{\rule{0.717pt}{0.400pt}}
\multiput(384.00,382.95)(1.802,-0.447){3}{\rule{1.300pt}{0.108pt}}
\multiput(384.00,383.17)(6.302,-3.000){2}{\rule{0.650pt}{0.400pt}}
\multiput(393.00,379.94)(1.358,-0.468){5}{\rule{1.100pt}{0.113pt}}
\multiput(393.00,380.17)(7.717,-4.000){2}{\rule{0.550pt}{0.400pt}}
\multiput(403.00,375.94)(1.358,-0.468){5}{\rule{1.100pt}{0.113pt}}
\multiput(403.00,376.17)(7.717,-4.000){2}{\rule{0.550pt}{0.400pt}}
\multiput(413.00,371.94)(1.358,-0.468){5}{\rule{1.100pt}{0.113pt}}
\multiput(413.00,372.17)(7.717,-4.000){2}{\rule{0.550pt}{0.400pt}}
\multiput(423.00,367.93)(0.933,-0.477){7}{\rule{0.820pt}{0.115pt}}
\multiput(423.00,368.17)(7.298,-5.000){2}{\rule{0.410pt}{0.400pt}}
\multiput(432.00,362.93)(1.044,-0.477){7}{\rule{0.900pt}{0.115pt}}
\multiput(432.00,363.17)(8.132,-5.000){2}{\rule{0.450pt}{0.400pt}}
\multiput(442.00,357.93)(0.852,-0.482){9}{\rule{0.767pt}{0.116pt}}
\multiput(442.00,358.17)(8.409,-6.000){2}{\rule{0.383pt}{0.400pt}}
\multiput(452.00,351.93)(0.852,-0.482){9}{\rule{0.767pt}{0.116pt}}
\multiput(452.00,352.17)(8.409,-6.000){2}{\rule{0.383pt}{0.400pt}}
\multiput(462.00,345.93)(0.645,-0.485){11}{\rule{0.614pt}{0.117pt}}
\multiput(462.00,346.17)(7.725,-7.000){2}{\rule{0.307pt}{0.400pt}}
\multiput(471.00,338.93)(0.626,-0.488){13}{\rule{0.600pt}{0.117pt}}
\multiput(471.00,339.17)(8.755,-8.000){2}{\rule{0.300pt}{0.400pt}}
\multiput(481.00,330.93)(0.553,-0.489){15}{\rule{0.544pt}{0.118pt}}
\multiput(481.00,331.17)(8.870,-9.000){2}{\rule{0.272pt}{0.400pt}}
\multiput(491.00,321.93)(0.553,-0.489){15}{\rule{0.544pt}{0.118pt}}
\multiput(491.00,322.17)(8.870,-9.000){2}{\rule{0.272pt}{0.400pt}}
\multiput(501.59,311.74)(0.489,-0.553){15}{\rule{0.118pt}{0.544pt}}
\multiput(500.17,312.87)(9.000,-8.870){2}{\rule{0.400pt}{0.272pt}}
\multiput(510.58,301.59)(0.491,-0.600){17}{\rule{0.118pt}{0.580pt}}
\multiput(509.17,302.80)(10.000,-10.796){2}{\rule{0.400pt}{0.290pt}}
\multiput(520.58,289.59)(0.491,-0.600){17}{\rule{0.118pt}{0.580pt}}
\multiput(519.17,290.80)(10.000,-10.796){2}{\rule{0.400pt}{0.290pt}}
\multiput(530.58,277.26)(0.491,-0.704){17}{\rule{0.118pt}{0.660pt}}
\multiput(529.17,278.63)(10.000,-12.630){2}{\rule{0.400pt}{0.330pt}}
\multiput(540.59,262.82)(0.489,-0.844){15}{\rule{0.118pt}{0.767pt}}
\multiput(539.17,264.41)(9.000,-13.409){2}{\rule{0.400pt}{0.383pt}}
\multiput(549.58,247.93)(0.491,-0.808){17}{\rule{0.118pt}{0.740pt}}
\multiput(548.17,249.46)(10.000,-14.464){2}{\rule{0.400pt}{0.370pt}}
\multiput(559.58,231.60)(0.491,-0.912){17}{\rule{0.118pt}{0.820pt}}
\multiput(558.17,233.30)(10.000,-16.298){2}{\rule{0.400pt}{0.410pt}}
\multiput(569.58,213.26)(0.491,-1.017){17}{\rule{0.118pt}{0.900pt}}
\multiput(568.17,215.13)(10.000,-18.132){2}{\rule{0.400pt}{0.450pt}}
\multiput(579.59,192.34)(0.489,-1.310){15}{\rule{0.118pt}{1.122pt}}
\multiput(578.17,194.67)(9.000,-20.671){2}{\rule{0.400pt}{0.561pt}}
\multiput(588.58,169.60)(0.491,-1.225){17}{\rule{0.118pt}{1.060pt}}
\multiput(587.17,171.80)(10.000,-21.800){2}{\rule{0.400pt}{0.530pt}}
\multiput(598.58,145.10)(0.491,-1.381){17}{\rule{0.118pt}{1.180pt}}
\multiput(597.17,147.55)(10.000,-24.551){2}{\rule{0.400pt}{0.590pt}}
\put(266.0,405.0){\rule[-0.200pt]{2.409pt}{0.400pt}}
\end{picture}
\]
\caption{\emph{The wave function renormalization constant $Z^{\varphi }$
given by eq.~(\ref{Z}) as a function of $\nu $.}}
\label{f1}
\end{figure}
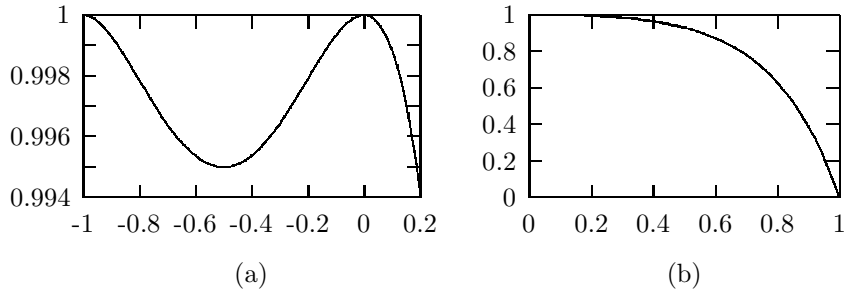

\begin{remark}
Usually in renormalized quantum field theory (in particular when $Z$ is
infinite) one would introduce the renormalized field with 
\[
\left\langle \,0\left| \varphi _{ren}(0)\right| p\right\rangle =1 
\]
by $\varphi (x)=\sqrt{Z^{\varphi }}\varphi _{ren}(x)$. Since Coleman's
article \cite{Co} however, the convention for the sine-Gordon model is to
use the unrenormalized field $\varphi (x)$ which is related to the massive
Thirring model current by 
\[
j^{\mu }=-\frac{\beta }{2\pi }\epsilon ^{\mu \nu }\partial _{\nu }\varphi
\,. 
\]
Therefore we have the normalization 
\[
\left\langle \,0\left| \varphi (0)\right| p\right\rangle =\sqrt{Z^{\varphi }}%
\,. 
\]
The wave function renormalization constant $Z^{\varphi }$ is plotted as a
function of $\nu $ in figure \ref{f1} (a) for negative values of $\nu $
which correspond to the sinh-Gordon model and in (b) for $0\leq \nu \leq 1$
which correspond to the sine-Gordon model for $0\leq \beta ^{2}\leq 4\pi $.
Note that in (a) the function is symmetric with respect to the self-dual
point $\nu =1/2$ of the sinh-Gordon model and that in (b) $Z^{\varphi }=1$
for the free breather point $\nu =0$ and $Z^{\varphi }=0$ for the free fermi
point $\nu =1$ where the breather disappears from the particle spectrum.
\end{remark}

As a generalization we now consider general n-particle form factors of an
normal ordered arbitrary power of the field $\mathcal{O}=\,:\!\varphi
^{N}\!: $ and let $m$ of the momenta tend to infinity. If $\underline{\theta 
}(\lambda )=(\theta _{1}+\lambda ,\dots ,\theta _{m}+\lambda ,\theta
_{m+1},\dots ,\theta _{n})\,,\,\underline{\theta }^{\prime }=(\theta
_{1},\dots ,\theta _{m})$ and $\underline{\theta }^{\prime \prime }=(\theta
_{m+1},\dots ,\theta _{n})$ Weinberg's power counting theorem for bosonic
Feynman graphs says that for $\mathop{\rm Re}\lambda \rightarrow \infty $%
\begin{eqnarray}
\left[ \varphi ^{N}\right] _{n}(\underline{\theta }(\lambda )) &\approx
&\sum_{K=0}^{N}\binom{N}{K}\left[ \varphi ^{K}\right] _{m}(\underline{\theta 
}^{\prime })\,\left[ \varphi ^{N-K}\right] _{n-m}(\underline{\theta }%
^{\prime \prime })  \label{3.8} \\
\begin{array}{c}
{\unitlength4.5mm%
\begin{picture}(6,6) \put(3,2){\oval(6,2)} \put(3,5){\makebox(0,0){$\bullet$}} \put(3,5.8){\makebox(0,0){${\cal
O}=\,:\!\varphi^N\!:$}} \put(1,3){\line(1,1){2}} \put(2,3){\line(1,2){1}} \put(5,3){\line(-1,1){2}} \put(3,.5){$\dots$} \put(2.7,3.5){$N$} \put(1,0){\line(0,1){1}} \put(2,0){\line(0,1){1}} \put(5,0){\line(0,1){1}} \put(-1.6,0){$\theta_1+\lambda$} \put(5.4,0){$\theta_n$} \end{picture} 
}
\end{array}
~ &\approx &\sum_{K=0}^{N}\binom{N}{K}~ 
\begin{array}{c}
\unitlength4.5mm\begin{picture}(11,6) \put(2,2){\oval(4,2)}
\put(8,2){\oval(4,2)} \put(5,5){\makebox(0,0){$\bullet$}}
\put(5,5.8){\makebox(0,0){${\cal O}=\,:\!\varphi^N\!:$}}
\put(1,3){\line(2,1){4}} \put(3,3){\line(1,1){2}} \put(7,3){\line(-1,1){2}}
\put(9,3){\line(-2,1){4}} \put(1.5,.5){$\dots$} \put(7.5,.5){$\dots$}
\put(1.7,4){$K$} \put(8,4){$N-K$} \put(1,0){\line(0,1){1}}
\put(3,0){\line(0,1){1}} \put(7,0){\line(0,1){1}} \put(9,0){\line(0,1){1}}
\put(-.1,0){$\theta_1$} \put(3.4,0){$\theta_m$} \put(9.4,0){$\theta_n$}
\end{picture}
\end{array}
\nonumber
\end{eqnarray}
with the notation $\left[ \varphi ^{N}\right] _{n}(\underline{\theta }%
)=\langle \,0\,|\!:\!\varphi ^{N}\!:\!(0)\,|\,p_{1},\dots ,p_{n}\,\rangle
^{in}$. For the special case of a local operator which is an exponential of
the fundamental bose field $\mathcal{O}=\,:e^{i\gamma \varphi }:$ (for some $%
\gamma )$ we therefore have 
\[
\left[ e^{i\gamma \varphi }\right] _{n}(\underline{\theta }(\lambda
))=\left[ e^{i\gamma \varphi }\right] _{m}(\underline{\theta }^{\prime
})\,\left[ e^{i\gamma \varphi }\right] _{n-m}(\underline{\theta }^{\prime
\prime })+O(e^{-\lambda })\,. 
\]
This gives in particular for $m=1$ and $\mathop{\rm Re}\theta
_{1}\rightarrow \infty $ 
\begin{equation}
\left[ e^{i\gamma \varphi }\right] _{n}(\theta _{1,}\theta _{2,}\dots
,\theta _{n})=\left[ e^{i\gamma \varphi }\right] _{1}(\theta _{1})\,\left[
e^{i\gamma \varphi }\right] _{n-1}(\theta _{2,}\dots ,\theta
_{n})+O(e^{-\theta _{1}})\,.  \label{3.9}
\end{equation}

\section{Examples of operators\label{s4}}

In this section we present some examples of p-functions which satisfy the
conditions of section \ref{s2} in particular (\ref{pp1}) -- (\ref{pp5}) and
propose the correspondence of local operators, K-functions and p-functions
due to eqs.~(\ref{1.1}) and (\ref{1.2}) for these examples 
\[
\mathcal{O}\leftrightarrow K_{n}^{\mathcal{O}}(\underline{\theta }%
)\,\leftrightarrow p_{n}^{\mathcal{O}}(\underline{\theta },\underline{l})\,. 
\]

\subsection{Classical local operators}

The classical sine-Gordon Lagrangian is 
\[
\mathcal{L}^{SG}=\tfrac{1}{2}\partial _{\mu }\varphi \partial ^{\mu }\varphi
+\frac{\alpha }{\beta ^{2}}\left( \cos \beta \varphi -1\right) 
\]
We consider the following classical local operators:

\begin{enumerate}
\item  $\exp \left( i\gamma \varphi (x)\right) $ for arbitrary real $\gamma
, $

\item  $\varphi ^{N}(x)$

\item  higher conserved currents for ($L=1,3,5.\dots $) 
\[
J_{L}^{\rho }=\left\{ 
\begin{array}{c}
J_{L}^{+}=\partial ^{+}\varphi \left( \partial ^{+}\right) ^{L}\varphi
+O(\varphi ^{4}) \\ 
J_{L}^{-}=\left( \left( \partial ^{+}\right) ^{L-1}\varphi +O(\varphi
^{2})\right) \sin \varphi
\end{array}
\right. . 
\]
A second set of conserved currents is obtained by replacing $\partial
^{+}\rightarrow \partial ^{-}$. The higher charges are of the form 
\[
Q_{L}=\int dx\left( \partial ^{0}\varphi \partial ^{+L}\varphi +O(\varphi
^{4})\right) \,,\quad L=1,3,5,\dots 
\]
and the charges for even $L$ vanish.

\item  $T^{\mu \nu }(x)$ $=\partial ^{\mu }\varphi \partial ^{\nu }\varphi
-g^{\mu \nu }\mathcal{L}^{SG}$ the energy momentum tensor or in terms of
light cone coordinates ($\partial ^{\pm }=\partial ^{0}\pm \partial ^{1}$
etc.) 
\begin{eqnarray*}
T^{\pm \pm }=T^{00}\pm T^{01}\pm T^{10}+T^{11}=\partial ^{+}\varphi \partial
^{+}\varphi =\partial ^{-}\varphi \partial ^{-}\varphi \\
T^{+-}=T^{00}-T^{01}+T^{10}-T^{11}=-2\frac{\alpha }{\beta ^{2}}\left( \cos
\beta \varphi -1\right) =T^{-+}
\end{eqnarray*}

\item  $\exp \left( i\beta \varphi (x)\right) $ for the particular value $%
\gamma =\beta $
\end{enumerate}

\subsection{The normalization of form factors}

The normalization constants are obtained in the various cases by the
following observations:

\begin{enumerate}
\item[a)]  The normalization a field annihilating a one-particle state is
given by the vacuum one-particle matrix element, in particular for the
fundamental bose field one has 
\begin{equation}
\langle \,0\,|\,\varphi (0)\,|\,p\,\rangle =\sqrt{Z^{\varphi }}\,.
\label{Na}
\end{equation}
$Z^{\varphi }$ is the finite wave function renormalization constant (\ref{Z}%
) which has been calculated in \cite{KW}.

\item[b)]  If an observable like a charge $Q=\int dx\,\mathcal{O}(x)$
belongs to a local operator we use the relation 
\[
\langle \,p^{\prime }\,|\,Q\,|\,p\,\rangle =q\langle \,p^{\prime
}\,|\,\,p\,\rangle . 
\]
This will be applied for example to the higher conserved charges.

\item[c)]  We use Weinberg's power counting theorem for bosonic Feynman
graphs. As discussed in section \ref{s3} this yields in particular the
asymptotic behavior for the exponentials of the boson field $\mathcal{O}%
=\,:\!e^{i\gamma \varphi }\!:$ 
\[
\mathcal{O}_{n}(\theta _{1,}\theta _{2,}\dots )=\mathcal{O}_{1}(\theta
_{1})\,\mathcal{O}_{n-1}(\theta _{2,}\dots )+O(e^{-\mathop{\rm Re}\theta
_{1}}) 
\]
as $\mathop{\rm Re}\theta _{1}\rightarrow \infty $ in any order of
perturbation theory. This behavior is also assumed to hold for the exact
form factors. Applying this formula iteratively we obtain from (\ref{1.2})
relations\footnote{%
This type of arguments has been also used in \cite{KW,FMS,KM,MS}.} for the
normalization constants of the operators $:e^{i\gamma \varphi }:$

\item[d)]  The recursion relation (\ref{pp3}) relates $N_{n+2}$ and $N_{n}.$
For all p-functions discussed below this means 
\begin{equation}
N_{n+2}=N_{n}\frac{2}{\sin \pi \nu F(i\pi )}\quad (n\geq 1).  \label{N}
\end{equation}
where $F(i\pi )$ is related to the wave function renormalization constant by 
\[
\frac{1}{F(i\pi )}=\frac{\beta ^{2}}{8\pi \nu }\frac{\sin \pi \nu }{\pi \nu }%
Z^{\varphi } 
\]
see \cite{BK} and eq.~(\ref{Z}).
\end{enumerate}

\subsection{Local operators and their p-functions}

For all cases to be discussed in the following, the conditions (\ref{pp1}-%
\ref{pp5}) are again obvious except that of the recursion relation (\ref{pp3}%
) which will be discussed in detail. For later convenience we also list for
some cases the explicit expressions of $K_{n}(\underline{\theta })$ for $%
n=1,2$ and the asymptotic behavior of $K_{n}(\underline{\theta })$ for $%
\mathop{\rm Re}\theta _{1}\rightarrow \infty $ which is easily obtained
analogously to the calculation (\ref{as}) in the proof of lemma \ref{l1}.
For convenience we will use the notation $K_{n}=N_{n}\tilde{K}_{n}$ in the
following.

\subsubsection{Exponentials of the breather field}

We propose the correspondence 
\begin{equation}
e^{i\gamma \varphi }\leftrightarrow N_{n}^{(q)}\tilde{K}_{n}^{(q)}(%
\underline{\theta })\leftrightarrow p_{n}^{(q)}(\underline{l}%
)=N_{n}^{(q)}\prod\limits_{i=1}^{n}q^{(-1)^{l_{i}}}  \label{p1}
\end{equation}
with $q=q(\gamma )$ (and $q(0)=1$) to be determined below. For low particle
numbers one easily calculates the K-functions 
\begin{equation}
\begin{array}{l}
\tilde{K}_{1}^{(q)}(\theta )=\left( q-1/q\right) \\ 
\tilde{K}_{2}^{(q)}(\underline{\theta })=\left( q-1/q\right) ^{2}
\end{array}
\label{4.1}
\end{equation}
and the asymptotic behavior 
\[
\tilde{K}_{n}^{(q)}(\underline{\theta })\approx \tilde{K}_{1}^{(q)}(\theta
_{1})\tilde{K}_{n-1}^{(q)}(\underline{\theta }^{\prime })\,. 
\]
The last formula is obtained similarly to the proof of lemma \ref{l1}. The
proposal that the p-function $p_{n}^{(q)}(\underline{l})$ correspond to an
exponential of a bosonic field is supported by the following observation.
The asymptotic behavior is consistent with that of the form factors of
exponentials of bosonic fields (\ref{3.9}) as discussed in section \ref{s3}.
Indeed it reads in terms of the K-functions as $K_{n}^{(q)}(\underline{%
\theta })\approx K_{1}^{(q)}(\theta _{1})K_{n-1}^{(q)}(\underline{\theta }%
^{\prime })$ (since $F(\infty )=1$) provided the normalization constants
satisfy 
\[
N_{n}^{(q)}=N_{1}^{(q)}N_{n-1}^{(q)}\quad \Rightarrow \quad
N_{n}^{(q)}=\left( N_{1}^{(q)}\right) ^{n}. 
\]
This is what we discussed above under point c) to determine the
normalization constants. Point d) in the present case has the following
meaning. The recursion condition (\ref{pp3}) is satisfied since in this case
we have $%
g(l_{1},l_{2})=q^{(-1)^{l_{1}}+(-1)^{l_{2}}}N_{n}^{(q)}/N_{n-2}^{(q)}$ which
is symmetric and $h(l_{1},l_{2})=0$. The condition (\ref{pp3}) with $%
g(0,1)=g(1,0)=2/(F(i\pi )\sin \pi \nu )$ implies the recursion relation for
the normalization constants (\ref{N}) which finally yields 
\begin{eqnarray}
N_{1}^{(q)} &=&\sqrt{\frac{2}{F(i\pi )\sin \pi \nu }}=\sqrt{Z^{\varphi }}%
\frac{\beta }{2\pi \nu }  \nonumber \\
N_{n}^{(q)} &=&\left( \sqrt{Z^{\varphi }}\frac{\beta }{2\pi \nu }\right) ^{n}
\label{Nq}
\end{eqnarray}
where $Z^{\varphi }$ is the breather wave function renormalization constant (%
\ref{Z}). The relation of $F(i\pi )$ with $Z^{\varphi }$ is obtained by
elementary manipulations of the integral representation (\ref{2.1}) and (\ref
{Z}). Recall that normal ordering implies $N_{0}^{(q)}=1$.

\subsubsection{Powers of the breather field}

Motivated by the expansion of (\ref{p1}) with respect to $\ln q$ we propose
the correspondence 
\begin{equation}
\varphi ^{N}\leftrightarrow N_{n}^{(N)}\tilde{K}_{n}^{(N)}(\underline{\theta 
})\leftrightarrow p_{n}^{(N)}(\underline{l})=N_{n}^{(N)}\left(
\sum\limits_{i=1}^{n}(-1)^{l_{i}}\right) ^{N}  \label{p2}
\end{equation}
Again one easily calculates (with $\tilde{K}%
_{n}^{(N)}=K_{n}^{(N)}/N_{n}^{(N)})$ the low particle number K-functions 
\[
\begin{array}{l}
\tilde{K}_{1}^{(N)}(\theta )=2 \\ 
\tilde{K}_{2}^{(N)}(\underline{\theta })=2^{N+1} \\ 
\tilde{K}_{3}^{(N)}(\underline{\theta })=2(3^{N}-3)-{\sin ^{2}\pi \nu }%
\prod\limits_{i<j}\dfrac{1}{\cosh \frac{1}{2}\theta _{ij}}
\end{array}
\]
and the asymptotic behavior 
\[
\tilde{K}_{n}^{(N)}(\underline{\theta })\approx \sum_{K=1}^{N}\binom{N}{K}%
\tilde{K}_{1}^{(K)}(\theta _{1})\tilde{K}_{n-1}^{(N-K)}(\underline{\theta }%
^{\prime }) 
\]
where $K_{n}^{(N)}$ is only nonvanishing for $N-n=$ even. This asymptotic
behavior agrees with (\ref{3.8}) which follows from Weinberg's power
counting argument and therefore justifies the correspondence (\ref{p2}). The
normalization condition $\langle \,0\,|\,\varphi (0)\,|\,p\,\rangle =\sqrt{%
Z^{\varphi }}$ ( see (\ref{Na})) yields 
\[
N_{1}^{(1)}=\tfrac{1}{2}\sqrt{Z^{\varphi }}\,. 
\]
The other normalization constants and also the function $q(\gamma )$ is now
obtained as follows. Comparing the correspondences (\ref{p1}) and (\ref{p2})
we conclude 
\[
N_{n}^{(N)}=N_{n}^{(q)}\left( \frac{\ln q}{i\gamma }\right) ^{N}. 
\]
This implies for $N=n=1$ together with (\ref{Nq}) 
\[
q=\exp \left( i\gamma \frac{N_{1}^{(1)}}{N_{1}^{(q)}}\right) =\exp \left( i%
\frac{\pi \nu }{\beta }\gamma \right) . 
\]
and finally the normalization constants 
\begin{equation}
N_{n}^{(N)}=\left( \frac{1}{2}\sqrt{Z^{\varphi }}\right) ^{n}\left( \frac{%
\pi \nu }{\beta }\right) ^{N-n}.  \label{4.2}
\end{equation}
We compare these general results with known special cases \cite{KW}. In
particular for $n=N=2$ we have 
\begin{eqnarray*}
\langle \,0\,|\!:\!\varphi ^{2}\!:\!(0)\,|\,p_{1},p_{2}\,\rangle ^{in}
&=&K_{2}^{(2)}(\theta _{12})F(\theta _{12})=8N_{2}^{(2)}F(\theta _{12}) \\
&=&2Z^{\varphi }F(\theta _{12})
\end{eqnarray*}
which agrees with formulae (4.4--4.6) of \cite{KW}. Further for $n=3$ and $%
N=1$ we have 
\begin{eqnarray*}
\langle \,0\,|\,\varphi (0)\,|\,p_{1},p_{2}\,,p_{3}\,\rangle ^{in}
&=&K_{3}^{(1)}(\underline{\theta })\prod_{i<j}F(\theta _{ij}) \\
&=&-\left( Z^{\varphi }\right) ^{3/2}\frac{1}{8}\left( \beta \frac{\sin \pi
\nu }{\pi \nu }\right) ^{2}\prod_{i<j}\frac{F(\theta _{ij})}{\cosh \frac{1}{2%
}\theta _{ij}}
\end{eqnarray*}
which again agrees with formulae (4.9--4.12) of \cite{KW}.

\subsubsection{Higher conserved currents}

In the following we present new results concerning the higher conservation
laws which are typical for integrable quantum field theories\footnote{\label%
{c2}In \cite{Sm1} form factors of higher currents in the sine-Gordon model
were proposed, however the charges of these currents vanish. The densities
where proposed to be of the form $\left( \partial ^{+}\right) ^{L}\partial
^{1}A(x)$ where the operator $A$ is related to the energy momentum tensor,
in particular $\partial ^{0}\partial ^{1}A=T^{10}$ and $\partial
^{1}\partial ^{1}A=T^{00}$. For $L>1$ obviously $\int dx\left( \partial
^{+}\right) ^{L}\partial ^{1}A(x)\allowbreak =\int dx\left( \partial
^{+}\right) ^{L-2}\partial ^{1}\left( T^{10}+T^{00}+T^{01}+T^{00}\right)
\allowbreak =0$ where the conversation laws $\partial ^{0}T^{\mu 0}=\partial
^{1}T^{\mu 1}$ have been used.}. We propose the correspondence 
\[
J_{L}^{\pm }\leftrightarrow p_{n}^{(L,\pm )}(\underline{\theta },\underline{l%
})=\pm N_{n}^{(J_{L})}\sum_{i=1}^{n}e^{\pm \theta
_{i}}\sum_{i=1}^{n}e^{L\left( \theta _{i}-\frac{i\pi }{2}(1-(-1)^{l_{i}}\nu
)\right) } 
\]
for $n=$ even and zero for $n=$ odd ($L=\pm 1,\pm 3,\dots $). Again one
easily calculates the 2 particle K-function 
\begin{equation}
\tilde{K}_{2}^{(L,\pm )}(\underline{\theta })=-2\left( -i\right) ^{L}\sin 
\tfrac{1}{2}L\pi \nu \dfrac{\sin \pi \nu }{\sinh \theta _{12}}\left( e^{\pm
\theta _{1}}+e^{\pm \theta _{2}}\right) \left( e^{L\theta _{1}}-e^{L\theta
_{2}}\right)  \label{4.0}
\end{equation}
The recursion condition (\ref{pp3}) is satisfied since $%
g(l_{1},l_{2})=N_{n}^{(J_{L})}/N_{n-2}^{(J_{L})}$ is symmetric and $%
h(l_{1},l_{2})=\sum_{i=3}^{n}e^{\pm \theta _{i}}\sum_{i=1}^{2}e^{L\left(
\theta _{i}-\frac{i\pi }{2}(1-(-1)^{l_{i}}\nu )\right) }$ is independent of $%
l_{i},\,(i>2)$. Again we have the recursion relation for the normalization
constants (\ref{N}). The two particle normalization we calculated by means
of b) with the charges 
\begin{align*}
\langle \,p^{\prime }\,|Q_{L}\,|\,p\rangle & =\int_{-\infty }^{\infty
}dx\,\langle \,p^{\prime }\,|\tfrac{1}{2}\left(
J_{L}^{+}(x)+J_{L}^{-}(x)\right) \,|\,p\rangle \\
& =\int_{-\infty }^{\infty }dx\,e^{i(p-p^{\prime })x}\,\tfrac{1}{2}\left(
K_{2}^{(L,+)}+K_{2}^{(L,-)}\right) (\theta ^{\prime }+i\pi ,\theta )F(\theta
^{\prime }+i\pi -\theta ) \\
& =2\pi \delta (p-p^{\prime })\tfrac{1}{2}\left(
K_{2}^{(L,+)}+K_{2}^{(L,-)}\right) (\theta +i\pi ,\theta )F(i\pi ) \\
& =\langle \,p^{\prime }\,|\,p\rangle \,e^{L\theta }\quad \text{if }L\;\text{%
odd.}
\end{align*}
Using (\ref{4.0}) we obtain 
\begin{align*}
& \tfrac{1}{2}\left( K_{2}^{(L,+)}+K_{2}^{(L,-)}\right) (\theta +i\pi
,\theta ) \\
& =-N_{n}^{(J_{L})}2\left( -i\right) ^{L}\sin \tfrac{1}{2}L\pi \nu \sin \pi
\nu \cosh \theta e^{L\theta }\left( e^{Li\pi }-1\right) \\
& =2m\cosh \theta \,e^{L\theta }/F(i\pi )
\end{align*}
for $L$ odd. For even $L$ the charges vanish as in the classical case. With
the relation of the normalization constants (\ref{N}) we finally obtain 
\begin{equation}
N_{n}^{(J_{L})}=\frac{m\,i^{L}}{4\sin \tfrac{1}{2}L\pi \nu }\left( \sqrt{%
Z^{\varphi }}\frac{\beta }{2\pi \nu }\right) ^{n}.  \label{Nj}
\end{equation}

Next we derive all eigenvalues of the higher charges (\ref{J}). We show that
for $n^{\prime }+n>2$ the connected part of the matrix element $%
^{out}\langle \,p_{1}^{\prime },\dots ,p_{n^{\prime }}^{\prime
}\,|Q_{L}|\,p_{1},\dots ,p_{n}\rangle ^{in}$ vanishes. The analytic
continuation $\mathcal{O}_{n^{\prime }+n}(\underline{\theta ^{\prime }}+i\pi
,\underline{\theta })$ yields this connected part. From the correspondence
of operators and p-functions 
\begin{multline*}
Q_{L}=\int dxJ_{L}^{0}(x)\leftrightarrow 2\pi \delta (P^{\prime
}-P)N_{n}^{(J_{L})}m\left( -\sum_{i=1}^{n^{\prime }}\sinh \theta
_{i}^{\prime }+\sum_{i=1}^{n}\sinh \theta _{i}\right) \\
\times \left( \sum_{i=1}^{n^{\prime }}e^{L\left( \theta _{i}^{\prime }+i\pi -%
\frac{i\pi }{2}(1-(-1)^{l_{i}}\nu )\right) }+\sum_{i=1}^{n}e^{L\left( \theta
_{i}-\frac{i\pi }{2}(1-(-1)^{l_{i}}\nu )\right) }\right)
\end{multline*}
the claim follows since for $n^{\prime }+n>2$ there are no poles which may
cancel the zero at $P^{\prime }=P$ where $P^{(\prime )}=\sum p_{i}^{(\prime
)}$. Therefore contributions to matrix element come from disconnected parts
which contain (analytic continued) two-particle form factors: 
\begin{align*}
& ^{out}\langle \,p_{1}^{\prime },\dots ,p_{n^{\prime }}^{\prime
}\,|Q_{L}\,|\,p_{1},\dots ,p_{n}\rangle ^{in} \\
& =\sum_{i,j}\,^{out}\langle \,p_{1}^{\prime },\dots ,\hat{p}_{i}^{\prime
},\dots ,p_{n^{\prime }}^{\prime }\,|\,p_{1},\dots ,\hat{p}_{j},\dots
,p_{n}\rangle ^{in}\,\langle \,p_{i}^{\prime }\,|Q_{L}\,|\,p_{j}\rangle \\
& =\,^{out}\langle \,p_{1}^{\prime },\dots ,p_{n^{\prime }}^{\prime
}\,|\,p_{1},\dots ,p_{n}\rangle ^{in}\,\sum_{i=1}^{n}e^{\theta _{i}L}
\end{align*}
where $\hat{p}_{j}$ means that this particles is missing in the state.

From the higher currents for $L=\pm 1$ we get the light cone components of
the energy momentum tensor $T^{\rho \sigma }\varpropto J_{\sigma }^{\rho }$
with $\rho ,\sigma =\pm $ (see also \cite{MS}).

\subsubsection{The energy momentum tensor}

We propose the correspondence 
\begin{equation}
T^{\rho \sigma }\leftrightarrow p_{n}^{\rho \sigma }(\underline{\theta },%
\underline{l})=\rho N_{n}^{(T)}\sum\limits_{i=1}^{n}e^{\rho \theta
_{i}}\sum\limits_{i=1}^{n}e^{\sigma \left( \theta _{i}-\frac{i\pi }{2}%
(1-(-1)^{l_{i}}\nu )\right) }  \label{p4}
\end{equation}
for $n$ even and $p_{n}^{\rho \sigma }=0$ for $n$ odd. The normalization is
again determined by c) namely 
\begin{equation}
\langle \,p^{\prime }\,|\,P^{\nu }\,|\,p\rangle =\langle \,p^{\prime
}\,|\,\int dxT^{0\nu }(x)\,|\,p\rangle =\langle \,p^{\prime }\,|\,p\rangle
\,p^{\nu }  \label{Nt}
\end{equation}
which in analogy to (\ref{Nj}) gives 
\begin{equation}
N_{n}^{(T)}=\frac{i\,m^{2}}{4\sin \tfrac{1}{2}\pi \nu }\left( \sqrt{%
Z^{\varphi }}\frac{\beta }{2\pi \nu }\right) ^{n}.  \label{4.4}
\end{equation}
Note that at the first sight the energy momentum tensor does not seem to be
symmetric. However, it is due to an identity proven in the next section (see
theorem \ref{t4}). The conservation law follows as above for the higher
currents and also the eigenvalue equation of the energy momentum operator
with the result 
\[
\,P^{\nu }\,|\,p_{1},\dots ,p_{n}\rangle ^{in}=\sum_{i=1}^{n}p_{i}^{\nu
}\,|\,p_{1},\dots ,p_{n}\rangle ^{in}. 
\]

\subsubsection{Special exponentials of the breather field}

For the special cases of the exponential of the field $\gamma =\pm \beta $
we propose the alternative correspondence to (\ref{p1}) 
\begin{equation}
e^{\pm i\beta \varphi }\leftrightarrow N_{n}^{\pm }\tilde{K}_{n}^{\pm }(%
\underline{\theta })\leftrightarrow p_{n}^{\pm }(\underline{\theta },%
\underline{l})=N_{n}^{\pm }\sum\limits_{i=1}^{n}e^{\mp \theta
_{i}}\sum\limits_{i=1}^{n}e^{\pm \left( \theta _{i}-\frac{i\pi }{2}%
(1-(-1)^{l_{i}}\nu )\right) }.  \label{p5}
\end{equation}
Again one easily calculates for low particle number the K-functions 
\[
\begin{array}{l}
\tilde{K}_{1}^{\pm }(\theta )=2\sin \tfrac{1}{2}\pi \nu \\ 
\tilde{K}_{2}^{\pm }(\underline{\theta })=\pm 4i\sin \tfrac{1}{2}\pi \nu
\sin \pi \nu
\end{array}
\]
and the asymptotic behavior 
\[
\tilde{K}_{n}^{\pm }(\underline{\theta })\approx \pm 2i\sin \pi \nu \,\tilde{%
K}_{n-1}^{\pm }(\underline{\theta }^{\prime })\,. 
\]
The proof of the last formula is delegated to the appendix. The
normalization constants are calculated analogously to the case of the
general exponential and take the form 
\begin{equation}
N_{n}^{\pm }=\pm i\frac{\sin \pi \nu }{\sin \tfrac{1}{2}\pi \nu }\left( 
\sqrt{Z^{\varphi }}\frac{\beta }{2\pi \nu }\right) ^{n}.  \label{4.3}
\end{equation}

\subsection{Identities\label{s4.4}}

It turns out that the correspondence between local operators and p-functions
is not unique. In this subsection we prove some identities which we will
need in the following section to prove operator equations. To have a
consistent interpretation of $K_{n}^{(q)}(\underline{\theta })$ with $q=\exp
\left( i\pi \nu \gamma /\beta \right) $ as the K-function of $\exp \left(
i\gamma \varphi (x)\right) $ it is necessary that $K_{n}^{(q)}(\underline{%
\theta })$ is even/odd for $n=$ even/odd under the exchange $%
q\leftrightarrow 1/q$. For $\gamma =\pm \beta $ the K-function of the
general exponential should turn into the K-functions of the special
exponentials. These fact are expressed by the following lemma.

\begin{lemma}
\label{l2}Let the K-functions 
\[
K_{n}(\underline{\theta })=\sum_{l_{1}=0}^{1}\dots
\sum_{l_{r}=0}^{1}(-1)^{l_{1}+\dots +l_{r}}\prod_{1\leq i<j\leq n}\left(
1+(l_{i}-l_{j})\frac{i\sin \pi \nu }{\sinh \theta _{ij}}\right) p_{n}(%
\underline{\theta },\underline{l}) 
\]
be given by the p-functions 
\begin{eqnarray*}
K_{n}^{(q)}(\underline{\theta })\leftrightarrow p_{n}^{(q)}(\underline{l}%
)=N_{n}^{(q)}\prod\limits_{i=1}^{n}q^{(-1)^{l_{i}}} \\
K_{n}^{\pm }(\underline{\theta })\leftrightarrow p_{n}^{\pm }(\underline{%
\theta },\underline{l})=N_{n}^{\pm }\sum\limits_{i=1}^{n}e^{\mp \theta
_{i}}\sum\limits_{i=1}^{n}e^{\pm \left( \theta _{i}-\frac{i\pi }{2}%
(1-(-1)^{l_{i}}\nu )\right) } \\
K_{n}^{(1)}(\underline{\theta })\leftrightarrow p_{n}^{(1)}(\underline{l}%
)=N_{n}^{(1)}\sum\limits_{i=1}^{n}(-1)^{l_{i}}\,.
\end{eqnarray*}
Then the following identities hold (again with $K_{n}(\underline{\theta }%
)=N_{n}\tilde{K}_{n}(\underline{\theta })$) 
\begin{eqnarray}
\tilde{K}_{n}^{(q)}(\underline{\theta })=-(-1)^{n}\tilde{K}_{n}^{(1/q)}(%
\underline{\theta })  \nonumber \\
\tilde{K}_{n}^{+}(\underline{\theta })=-(-1)^{n}\tilde{K}_{n}^{-}(\underline{%
\theta })\,,  \label{4.5}
\end{eqnarray}
in particular for $\gamma =\beta $ i.e. $q=\exp (i\pi \nu )$%
\[
K_{n}^{+}(\underline{\theta })=K_{n}^{(q)}(\underline{\theta }). 
\]
and further more 
\begin{equation}
\tilde{K}_{n}^{(1)}(\underline{\theta })=\frac{1}{2\sin \frac{1}{2}\pi \nu }%
\left( \sum_{i=1}^{n}e^{\theta _{i}}\sum_{i=1}^{n}e^{-\theta _{i}}\right)
^{-1}\left( \tilde{K}_{n}^{+}(\underline{\theta })+\tilde{K}_{n}^{-}(%
\underline{\theta })\right) .  \label{4.6}
\end{equation}
\end{lemma}

\proof  %
Again as in the proof of lemma \ref{l1} we use induction and Liouville's
theorem. We introduce the differences 
\begin{eqnarray*}
f_{n}(\underline{\theta }) &=&\tilde{K}_{n}^{(q)}(\underline{\theta }%
)+(-1)^{n}\tilde{K}_{n}^{(1/q)}(\underline{\theta }) \\
\text{or\quad }f_{n}(\underline{\theta }) &=&\tilde{K}_{n}^{+}(\underline{%
\theta })+(-1)^{n}\tilde{K}_{n}^{-}(\underline{\theta }) \\
\text{or\quad }f_{n}(\underline{\theta }) &=&K_{n}^{+}(\underline{\theta }%
)-K_{n}^{(q=\exp (i\pi \nu ))}(\underline{\theta })\, \\
\text{or\quad }f_{n}(\underline{\theta }) &=&\tilde{K}_{n}^{(1)}(\underline{%
\theta })-\frac{1}{2\sin \frac{1}{2}\pi \nu }\left( \sum_{i=1}^{n}e^{\theta
_{i}}\sum_{i=1}^{n}e^{-\theta _{i}}\right) ^{-1}\left( \tilde{K}_{n}^{+}(%
\underline{\theta })+\tilde{K}_{n}^{-}(\underline{\theta })\right) .
\end{eqnarray*}
Then the results of the previous subsection imply in all cases $f_{1}(\theta
)=f_{2}(\underline{\theta })=0$. As induction assumptions we take $f_{n-2}(%
\underline{\theta }^{\prime \prime })=0$. The functions $f_{n}(\underline{%
\theta })$ are meromorphic functions in terms of the $x_{i}=e^{\theta _{i}}$
with at most simple poles at $x_{i}=\pm x_{j}$ since $\sinh \theta
_{ij}=(x_{i}+x_{j})(x_{i}-x_{j})/(2x_{i}x_{j})$. The residues of the poles
at $x_{i}=x_{j}$ vanish because of the symmetry and again the residues at $%
x_{i}=-x_{j}$ are proportional to $f_{n-2}(\underline{\theta }^{\prime
\prime })$ due to the recursion relation (\ref{pk3}). Furthermore for $%
x_{i}\rightarrow \infty $ again $f_{n}(\underline{\theta })\rightarrow 0$.
Therefore $f_{n}(\underline{\theta })$ vanishes identically by Liouville's
theorem in for all $n$. For the last case of $f_{n}(\underline{\theta })$ it
has been used that because of (\ref{4.5}) for $n$ even both $\tilde{K}^{\pm
} $-terms cancel and that they are equal for odd $n$. Due to (\ref{p5}) $%
\tilde{K}_{n}^{+}$ is proportional to $\sum_{i=1}^{n}e^{-\theta _{i}}$ and $%
\tilde{K}_{n}^{-}$ is proportional to $\sum_{i=1}^{n}e^{\theta _{i}}$. Hence
both are proportional to $\sum_{i=1}^{n}e^{-\theta
_{i}}\sum_{i=1}^{n}e^{+\theta _{i}}$ which means that there are no extra
poles at $\sum_{i=1}^{n}e^{\theta _{i}}=0$ or $\sum_{i=1}^{n}e^{-\theta
_{i}}=0$.%
\endproof %

\section{Operator equations\label{s5}}

The classical sine-Gordon model is given by the wave equation 
\[
\square \varphi (t,x)+\frac{\alpha }{\beta }\sin \beta \varphi (t,x)=0. 
\]
If this is fulfilled we have also the relation for the trace of
energy-momentum tensor 
\[
T_{~\mu }^{\mu }(t,x)=-2\frac{\alpha }{\beta ^{2}}\left( \cos \beta \varphi
(t,x)-1\right) . 
\]
In this section we construct the quantum version of these two classical
equations. In the following $:\dots :$ denotes normal ordering with respect
to the physical vacuum which means in particular for the vacuum expectation
value $\langle \,0\,|\,:\exp i\gamma \varphi :(t,x)\,|\,\,0\,\rangle =1$. As
consequences of the identities of subsection \ref{s4.4} we can prove quantum
field equations.

\begin{theorem}
\label{t4}The following operator equations are to be understood in term of
all their matrix elements.

\begin{enumerate}
\item  For the exceptional value $\gamma =\beta $ the operator $\square
^{-1}:\sin \gamma \varphi :(t,x)$ is local and the quantum sine-Gordon field
equation holds\footnote{\label{c1}This field equation was also discussed in 
\cite{Sm1} and for the sinh-Gordon case in \cite{MS}. However, in these
articles the relations of the bare and the renormalized mass differs from (%
\ref{5.1}) and are not consistent with perturbation theory and the results
of \cite{Fa,AlZa}.} 
\begin{equation}
\square \varphi (t,x)+\frac{\alpha }{\beta }:\sin \beta \varphi :(t,x)=0
\label{e}
\end{equation}
if the ``bare'' mass $\sqrt{\alpha }$ is related to the renormalized one by%
\footnote{%
Before this formula was found in \cite{Fa,AlZa} by different methods.} 
\begin{equation}
\alpha =m^{2}\frac{\pi \nu }{\sin \pi \nu }\,.  \label{5.1}
\end{equation}
Here $m$ is the physical mass of the fundamental boson.

\item  The energy momentum tensor is symmetric and its trace satisfies%
\footnote{%
This equation was also obtained by Smirnov \cite{Sm1}.} 
\begin{equation}
T_{~\mu }^{\mu }(t,x)=-2\frac{\alpha }{\beta ^{2}}\left( 1-\frac{\beta ^{2}}{%
8\pi }\right) \left( :\cos \beta \varphi :(t,x)-1\right) .  \label{T}
\end{equation}

\item  For all higher currents the conservation laws hold 
\[
\partial _{\mu }J_{L}^{\mu }(x)=0\quad (L=\pm 1,\pm 3,\dots )\,.
\]
\end{enumerate}
\end{theorem}

\proof%

\begin{enumerate}
\item  From (\ref{p5}) we have the correspondence between operators and
K-functions 
\[
\square ^{-1}\sin \beta \varphi \,\leftrightarrow \frac{K_{n}^{+}(\underline{%
\theta })-K_{n}^{-}(\underline{\theta })}{2i\sum_{i=1}^{n}e^{\theta
_{i}}\sum_{i=1}^{n}e^{-\theta _{i}}}.
\]
As shown in the proof of lemma \ref{l2} there are no poles at $%
\sum_{i=1}^{n}e^{\theta _{i}}=0$ or $\sum_{i=1}^{n}e^{-\theta _{i}}=0$.
Therefore $\square ^{-1}:\sin \beta \varphi :$ is a local operator.
Furthermore by eq.~(\ref{4.6}) 
\[
\sum_{i=1}^{n}e^{\theta _{i}}\sum_{i=1}^{n}e^{-\theta _{i}}K_{n}^{(1)}(%
\underline{\theta })=\frac{\pi \nu }{\beta \sin \pi \nu }\frac{1}{2i}\left(
K_{n}^{+}(\underline{\theta })-K_{n}^{-}(\underline{\theta })\right) 
\]
where the normalizations (\ref{4.2}) and (\ref{4.3}) have been used. This
means in particular 
\[
\frac{N_{n}^{(1)}}{N_{n}^{+}}\frac{i}{\sin \frac{1}{2}\pi \nu }=\frac{\pi
\nu }{\beta \sin \pi \nu }\,.
\]
In terms of operators this is just the quantum sine-Gordon field equation.
Comparing this result with the classical equation we obtain the relation
eq.~(\ref{5.1}) between the bare and the physical mass.

\item  Using (\ref{p4}) and (\ref{p5}) we have the correspondence between
operators and K-functions for $n$ even 
\begin{eqnarray*}
T^{+-} &\leftrightarrow &N_{n}^{(T)}\tilde{K}_{n}^{-} \\
T^{-+} &\leftrightarrow &-N_{n}^{(T)}\tilde{K}_{n}^{+} \\
T_{~\mu }^{\mu } &\leftrightarrow &K_{n}^{(T)}(\underline{\theta })=-\frac{1%
}{2}N_{n}^{(T)}\left( \tilde{K}_{n}^{+}-\tilde{K}_{n}^{-}\right)  \\
\cos \beta \varphi -1\, &\leftrightarrow &\frac{1}{2}\left( K_{n}^{+}(%
\underline{\theta })+K_{n}^{-}(\underline{\theta })\right) 
\end{eqnarray*}
The symmetry $T^{+-}=T^{-+}$ is again a consequence of (\ref{4.5}).
Furthermore the identity of K-functions follows 
\[
K_{n}^{(T)}(\underline{\theta })=-\frac{\,\alpha \left( 1-\frac{\beta ^{2}}{%
8\pi }\right) }{\beta ^{2}}\left( K_{n}^{+}(\underline{\theta })+K_{n}^{-}(%
\underline{\theta })\right) 
\]
where the normalizations (\ref{4.4}) and (\ref{4.3}) have been used which
means 
\[
\frac{N_{n}^{(T)}}{N_{n}^{+}}=2\frac{\,\alpha }{\beta ^{2}}\left( 1-\frac{%
\beta ^{2}}{8\pi }\right) 
\]

\item  The claim follows since we have the correspondence of operators and
p-functions 
\begin{multline*}
\partial _{\mu }J_{L}^{\mu }\leftrightarrow P^{+}p_{n}^{(L,-)}(\underline{%
\theta },\underline{l})+P^{-}p_{n}^{(L,+)}(\underline{\theta },\underline{l}%
)=-N_{n}^{(J_{L})}m \\
\times \left( \sum_{i=1}^{n}e^{\theta _{i}}\sum_{i=1}^{n}e^{-\theta
_{i}}-\sum_{i=1}^{n}e^{-\theta _{i}}\sum_{i=1}^{n}e^{\theta _{i}}\right)
\sum_{i=1}^{n}e^{L\left( \theta _{i}-\frac{i\pi }{2}(1-(-1)^{l_{i}}\nu
)\right) }=0\,.
\end{multline*}
\endproof%
\end{enumerate}

The factor $\frac{\pi \nu }{\sin \pi \nu }$ modifies the classical equation
and has to be considered as a quantum correction. For the sinh-Gordon model
an analogous quantum field equation has been obtained in \cite{MS}\footnote{%
It should be obtained from (\ref{e}) by the replacement $\beta \rightarrow
ig $. However the relation between the bare and the renormalized mass in 
\cite{MS} differs from the analytic continuation of (\ref{5.1}).}. Note that
in particular at the `free fermion point' $\nu \rightarrow 1\;(\beta
^{2}\rightarrow 4\pi )$ this factor diverges, a phenomenon which is to be
expected from short distance investigations \cite{ST}. For fixed bare mass
square $\alpha $ and $\nu \rightarrow 2,3,4,\dots $ the physical mass goes
to zero. These values of the coupling are known to be special: 1. the Bethe
Ansatz vacuum in the language of the massive Thirring model shows phase
transitions \cite{Ko} and 2. the model at these points is related \cite
{K3,LeC,Sm2} to Baxters RSOS-models which correspond to minimal conformal
models with central charge $c=1-6/(\nu (\nu +1))$ (see also \cite{MS}).

The second formula (\ref{T}) is consistent with renormalization group
arguments \cite{Z,Ca}. In particular this means that $\beta ^{2}/4\pi $ is
the anomalous dimension of $\cos \beta \varphi $. Again this operator
equation is modified by a quantum correction $(1-\beta ^{2}/8\pi )$.
Obviously for fixed bare mass square $\alpha $ and $\beta ^{2}\rightarrow
8\pi $ the model will become conformal invariant. This in turn is related to
a Berezinski-Kosterlitz-Thouless phase transition \cite{KT,JKKN,Wi}. The
conservation law $\partial _{\mu }T^{\mu \nu }=0$ for the energy momentum
tensor holds, because it is obtained from the higher currents for $L=\pm 1$. 
\begin{figure}[tbh]
\[
\begin{array}{l}
\unitlength5mm \begin{picture}(14,3) \put(1,2){\oval(2,2)}
\put(0,2){\makebox(0,0){$\bullet$}} \put(2,2){\makebox(0,0){$\bullet$}}
\put(-.5,2){\line(1,0){3}} \put(2.8,2){$p^2=m^2$} \put(1.5,-.5){$(a)$}
\put(11,2){\oval(2,2)} \put(10,2){\makebox(0,0){$\bullet$}}
\put(12,2){\makebox(0,0){$\bullet$}} \put(12.4,2){$\cos \beta \varphi$}
\put(10,2){\line(-1,2){.5}} \put(10,2){\line(-1,-2){.5}} \put(8.8,1){$p$}
\put(8.8,2.7){$p$} \put(11,-.5){$(b)$} \end{picture}
\end{array}
\]
\caption{Feynman graphs}
\label{f8}
\end{figure}
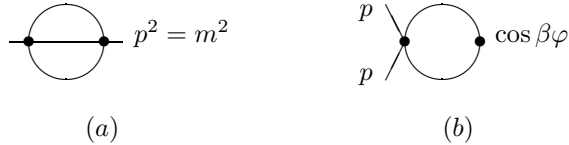
All the results may be checked in perturbation theory by Feynman graph
expansions. In particular in lowest order the relation between the bare and
the renormalized mass (\ref{5.1}) is given by figure \ref{f8} (a). It had
already been calculated in \cite{KW} and yields 
\[
m^{2}=\alpha \left( 1-\frac{1}{6}\left( \frac{\beta ^{2}}{8}\right)
^{2}+O(\beta ^{6})\right) 
\]
which agrees with the exact formula above. Similarly we check the quantum
corrections of the trace of the energy momentum tensor (\ref{T}) by
calculating the Feynman graph of Figure \ref{f8} (b) with the result again
taken from \cite{KW} as 
\[
\langle \,p\,|\!:\!\cos \beta \varphi -1\!:\!|\,p\,\rangle =-\beta
^{2}\left( 1+\frac{\beta ^{2}}{8\pi }\right) +O(\beta ^{6}). 
\]
This again agrees with the exact formula above since the normalization given
by eq.~(\ref{Nt}) implies $\langle \,p\,|\,T_{~\mu }^{\mu }|\,p\,\rangle
=2m^{2}$.

\section{Other representations of form factors\label{s6}}

\subsection{Determinant representation of bosonic sine-Gordon form factors}

The scaling Lee-Yang model is equivalent to the breather part of the
sine-Gordon model for the coupling constant equal to $\nu =1/3$ (in our
notation). For this model Smirnov \cite{Sm1} derived a determinant formula
for form factors (see also \cite{AlZa1}). Generalizing this formula in \cite
{FMS,KM} form factors were proposed for the sinh-Gordon model in terms of
determinants. The sinh-Gordon model form factors should be identified with
sine-Gordon form factors by analytic continuation $\nu \rightarrow $
negative values. Using this one would propose for the sine-Gordon model an
analogous determinant representation for the K-function of exponentials of
the field $\mathcal{O}=\,:e^{ik\beta \varphi }:$ 
\[
\tilde{K}_{n}^{(q)}(\underline{\theta })=\left( q^{\prime }-1/q^{\prime
}\right) ^{n}\prod_{i,j=1}^{n}(x_{i}+x_{j})^{-1}Det_{n}(\underline{x},k) 
\]
where $\underline{x}=(x_{1},\dots ,x_{n})$ with $x_{i}=e^{\theta _{i}}$ and $%
q^{\prime }=\exp i\pi \nu $. The Determinant is 
\begin{eqnarray*}
Det_{n}(\underline{x},k) &=&\det \left( \left( \left( k+i-j\right)
_{q^{\prime }}\sigma _{2i-j-1}(\underline{x})\right) _{i,j=1}^{n}\right) \\
&=&\left| 
\begin{array}{ccc}
\left( k\right) _{q^{\prime }}\sigma _{0} & \cdots & \left( k-n+1\right)
_{q^{\prime }}\sigma _{-n+1} \\ 
\vdots & \ddots & \vdots \\ 
\left( k+n-1\right) _{q^{\prime }}\sigma _{2n-2} & \cdots & \left( k\right)
_{q^{\prime }}\sigma _{n-1}
\end{array}
\right| .
\end{eqnarray*}
where the symmetric polynomials $\sigma _{l}(\underline{x})$ are defined by 
\[
\prod_{l=1}^{n}(y+x_{l})=\sum_{\lambda =0}^{n}y^{n-\lambda }\sigma _{\lambda
}^{(n)}(\underline{x})\, 
\]
and $\left( k\right) _{q^{\prime }}=\sin k\pi \nu /\sin \pi \nu $. This
relation of $\tilde{K}_{n}^{(q)}(\underline{\theta })$ with $Det_{n}(%
\underline{x},k)$ could be proven similar as in the proof of lemma \ref{l2},
once the corresponding recursion relation (\ref{pk3}) has been proven for $%
Det_{n}(\underline{x},k)$. This has been done \cite{FMS} only for the
special values $\nu =-1/2,-1/3$ of the coupling constant.

\subsection{Integral representations of breather form factors\label{s6.2}}

In \cite{BFKZ} and \cite{BK} integral representations for solitonic form
factors were proposed. These formulae are quite general and model
independent, so analogous formulae should also hold for breather form
factors. We propose for $n$ lowest breathers and $0\leq m\leq n$ 
\begin{equation}
\mathcal{O}_{n}(\underline{\theta })=\int_{\mathcal{C}_{\underline{\theta }%
}}dz_{1}\cdots \int_{\mathcal{C}_{\underline{\theta }}}dz_{m}\,h(\underline{%
\theta },{\underline{z}})\,\check{p}^{\mathcal{O}}(\underline{\theta },{%
\underline{z}})  \label{5.10}
\end{equation}
with the scalar function (c.f. \cite{BFKZ}) 
\[
h(\underline{\theta },{\underline{z}})=\prod_{1\le i<j\le n}F(\theta
_{ij})\prod_{i=1}^{n}\prod_{j=1}^{m}\tilde{\phi}(\theta
_{i}-z_{j})\prod_{1\le i<j\le m}\tau (z_{i}-z_{j}) 
\]
and 
\[
\tilde{\phi}(z)=\frac{S(z)}{F(z)\,F(z+i\pi )}=1+\frac{i\sin \pi \nu }{\sinh z%
}=\frac{\sinh z+i\sin \pi \nu }{\sinh z}\,. 
\]
\[
\tau (z)=\frac{1}{\tilde{\phi}(z)\,\tilde{\phi}(-z)}=\frac{\sinh ^{2}z}{%
\sinh ^{2}z+\sin ^{2}\pi \nu }\,. 
\]
The two breather form factor function $F(\theta )$ is again defined by eq.~(%
\ref{2.1}). For all integration variables $z_{j}$ $(j=1,\dots ,m)$ the
integration contours $\mathcal{C}_{\underline{\theta }}$ encloses clock wise
oriented the points $z_{j}=\theta _{i}\,(i=1,\dots ,n)$. The above integral
representation satisfies all form factor properties if suitable conditions
for the new type of p-function\footnote{%
A similar function was used by Cardy and Mussardo \cite{CM} in case of the
scaling Ising model to represent the various operators.} $\check{p}^{%
\mathcal{O}}(\theta ,z)$ are assumed. Here we consider $\check{p}=$
constant. The form factors of the exponential of the field $\mathcal{O}%
(x)=\,:\!\exp i\gamma \varphi \!:\!(x)$ are given by linear combinations of
expressions (\ref{5.10}) 
\begin{equation}
\mathcal{O}_{n}(\underline{\theta })=\left( \sqrt{Z^{\varphi }}\frac{\beta }{%
2\pi \nu }\right) ^{n}\prod_{1\leq i<j\leq n}F(\theta
_{ij})\sum_{m=0}^{n}q^{n-2m}(-1)^{m}I_{nm}(\underline{\theta })  \label{5.11}
\end{equation}
where again $q=\exp \left( i\frac{\pi \nu }{\beta }\gamma \right) $ and 
\begin{eqnarray*}
I_{nm}(\underline{\theta }) &=&\frac{1}{m!}\int_{\mathcal{C}_{\underline{%
\theta }}}\frac{dz_{1}}{R}\cdots \int_{\mathcal{C}_{\underline{\theta }}}%
\frac{dz_{m}}{R}\,\prod_{i=1}^{n}\prod_{j=1}^{m}\tilde{\phi}(\theta
_{i}-z_{j})\prod_{1\le i<j\le m}\tau (z_{i}-z_{j}) \\
&=&\sum_{\mathrel{\mathop{K\subseteq N}\limits_{|K|=m}}}\prod_{i\in
N\setminus K}\prod_{k\in K}\tilde{\phi}(\theta _{ik})
\end{eqnarray*}
with $R=\mathop{\rm Res}\limits_{\theta =0}\tilde{\phi}(\theta )$ and $%
N=\left\{ 1,\dots ,n\right\} ,K=\left\{ k_{1},\dots ,k_{m}\right\} $. It is
easy to verify that the asymptotic behavior (\ref{3.9}) is satisfied. Also
the low particle number form factors agree with eqs.~(\ref{4.1}). This
proves formula (\ref{5.11}).

Another integral representation is directly obtained from the integral
representations for solitonic form factors in \cite{BK} 
\begin{multline}
\mathcal{O}(\underline{\theta })=\tilde{N}_{n}^{\mathcal{O}%
}\prod_{i<j}\left( F(\theta _{ij})\tanh \tfrac{1}{2}\theta _{ij}\sinh \tfrac{%
1}{2}(\theta _{ij}+i\pi \nu )\sinh \tfrac{1}{2}(\theta _{ij}-i\pi \nu
)\right) \\
\times \int_{\mathcal{D}_{\theta _{1}}}dz_{1}\cdots \int_{\mathcal{D}%
_{\theta _{n}}}dz_{r}\prod_{i=1}^{n}\prod_{j=1}^{n}\chi (\theta
_{i}-z_{j})\prod_{i<j}\sinh z_{ij}\,\,p(\underline{\theta },\underline{z})\,
\label{6.6}
\end{multline}
where the contour $\mathcal{D}_{\theta _{i}}$ consists of two circles around
the poles at $\theta _{i}-\frac{i\pi }{2}(1\pm \nu )$ and 
\[
\chi (\theta )=\frac{1}{\sinh \frac{1}{2}(\theta -\frac{i\pi }{2}(1-\nu
))\sinh \frac{1}{2}(\theta -\frac{i\pi }{2}(1+\nu ))}\,. 
\]
As a matter of fact in \cite{BK} from this integral representation the
representation (\ref{1.1}) with the K-function (\ref{1.2}) was derived using
the identity 
\begin{multline*}
\sinh \tfrac{1}{2}(\theta _{ij}-i\pi \nu )\chi (\theta
_{i}-z_{j}^{(l_{j})})\chi (\theta _{j}-z_{i}^{(l_{i})})\sinh
(z_{i}^{(l_{i})}-z_{j}^{(l_{j})}) \\
=\frac{2}{\tanh \tfrac{1}{2}\xi _{ij}\sinh \tfrac{1}{2}(\xi _{ij}+i\pi \nu
)\sinh \tfrac{1}{2}(\xi _{ij}-i\pi \nu )}\left( 1+(l_{i}-l_{j})\frac{i\sin
\pi \nu }{\sinh \xi _{ij}}\right)
\end{multline*}
for $l_{i},l_{j}=0,1$ and $z_{i}^{(l_{i})}=\xi _{i}-\frac{i\pi }{2}%
(1-(-1)^{l_{i}}\nu )$. Performing one integration in eq.~(\ref{6.6}) and
using symmetry properties of the integrand one obtains an integral
representations of the type as used by Smirnov in \cite{Sm1}.

\paragraph{Acknowledgments:}

We thank J. Balog, A.A. Belavin, V.A. Fateev, R. Flu\-me, A. Fring, S.
Pakuliak, R.H. Poghossian, R. Schrader, B. Schroer, F.A. Smirnov and Al.B.
Zamolodchikov for discussions. One of authors (M.K.) thanks E. Seiler and P.
Weisz for discussions and hospitality at the Max-Planck Insitut f\"{u}r
Physik (M\"{u}nchen), where parts of this work have been performed. H.B. was
supported by DFG, Sonderforschungsbereich 288 `Differentialgeometrie und
Quantenphysik' and partially by grants INTAS 99-01459 and INTAS 00-561.

\appendix

\section*{Appendix: Asymptotic behavior}

\begin{lemma}
\label{l4.1}The K-functions defined by eq.~(\ref{1.2}) and the p-functions 
\[
\begin{array}{lll}
\mathrm{a)} &  & \tilde{p}_{n}^{(q)}(\underline{\theta })=\prod%
\limits_{i=1}^{n}q^{(-1)^{l_{i}}} \\ 
\mathrm{b)} &  & \tilde{p}_{n}^{(N)}(\underline{\theta })=\left(
\sum\limits_{i=1}^{n}(-1)^{l_{i}}\right) ^{N} \\ 
\mathrm{c)} &  & \tilde{p}_{n}^{(\pm )}(\underline{\theta }%
)=\sum\limits_{i=1}^{n}e^{\mp \theta _{i}}\sum\limits_{i=1}^{n}e^{\pm
z_{i}^{(l_{i})}}
\end{array}
\]
satisfy for $\mathop{\rm Re}\theta _{1}\rightarrow \infty $ the asymptotic
behavior 
\[
\begin{array}{lll}
\mathrm{a)} &  & \tilde{K}_{n}^{(q)}(\underline{\theta })=\tilde{K}%
_{1}^{(q)}(\theta _{1})\tilde{K}_{n-1}^{(q)}(\underline{\theta }^{\prime
})+O(e^{-\mathop{\rm Re}\theta _{1}}) \\ 
\mathrm{b)} &  & \tilde{K}_{n}^{(N)}(\underline{\theta })=\sum%
\limits_{K=1}^{N-1}\binom{N}{K}\tilde{K}_{1}^{(K)}(\theta _{1})\tilde{K}%
_{n-1}^{(N-K)}(\underline{\theta }^{\prime })+O(e^{-\mathop{\rm Re}\theta
_{1}}) \\ 
\mathrm{c)} &  & \tilde{K}_{n}^{\pm }(\underline{\theta })=\pm 2i\sin \pi
\nu \,\tilde{K}_{n-1}^{\pm }(\underline{\theta }^{\prime })+O(e^{-%
\mathop{\rm Re}\theta _{1}})
\end{array}
\]
with $\underline{\theta }^{\prime }=(\theta _{2},\dots ,\theta _{1})$. In
particular 
\[
\tilde{K}_{1}^{(1)}(\theta )=const\quad \text{and}\quad \tilde{K}_{n}^{(1)}(%
\underline{\theta })=O(e^{-\mathop{\rm Re}\theta _{1}})\quad \text{ for }%
n>1. 
\]
\end{lemma}

\proof %
The two first asymptotic relations are quite obvious. Note that $\tilde{K}%
_{1}^{(q)}=q-1/q$ and $\tilde{K}_{1}^{(1)}(\underline{\theta })=2$.

\begin{description}
\item[a)]  For $\tilde{K}_{n}^{(q)}(\underline{\theta })$ and $\mathop%
\mathrm{Re}\theta _{1}\rightarrow \infty $ we have 
\[
\tilde{K}_{n}^{(q)}(\underline{\theta })=%
\sum_{l_{1}=0}^{1}(-1)^{l_{1}}q^{(-1)^{l_{1}}}\tilde{K}_{n-1}^{(q)}(%
\underline{\theta }^{\prime })+O(e^{-\mathop{\rm Re}\theta _{1}}) 
\]

\item[b)]  For $\tilde{K}_{n}^{(N)}(\underline{\theta })$ we use 
\[
\left( \sum_{i=1}^{n}(-1)^{l_{i}}\right) ^{N}=\sum_{K=0}^{N}\binom{N}{K}%
\left( (-1)^{l_{1}}\right) ^{K}\left( \sum_{i=2}^{n}(-1)^{l_{i}}\right)
^{N-K} 
\]
which proves the claim as in the previous case.

\item[c)]  For $\tilde{K}_{n}^{+}(\underline{\theta })$ and $\mathop{\rm Re}%
\theta _{1}\rightarrow \infty $ we have 
\begin{align*}
\sum_{i=1}^{n}e^{-\theta _{i}}\sum_{i=1}^{n}e^{z_{i}^{(l_{i})}}=\left(
e^{-\theta _{1}}+\sum_{i=2}^{n}e^{-\theta _{i}}\right) \left( e^{\theta _{1}-%
\frac{i\pi }{2}(1-(-1)^{l_{1}}\nu )}+\sum_{i=2}^{n}e^{z_{i}^{(l_{i})}}\right)
\\
=e^{-\frac{i\pi }{2}(1-(-1)^{l_{1}}\nu )}+\left( \sum_{i=2}^{n}e^{-\theta
_{i}}\right) e^{\theta _{1}-\frac{i\pi }{2}(1-(-1)^{l_{1}}\nu )} \\
+\sum_{i=2}^{n}e^{-\theta _{i}}\sum_{i=2}^{n}e^{z_{i}^{(l_{i})}}+O(e^{-%
\mathop{\rm Re}\theta _{1}}).
\end{align*}
\end{description}

We calculate the leading terms $O(1).$ The contribution of the first term
consists of two types: one vanishes because of the lemma above and the other
is of order $O(e^{-\mathop{\rm Re}\theta _{1}})$. The contribution of the
third term vanishes after summation over $l_{1}.$ The contribution of the
second term is proportional to 
\begin{multline*}
\sum_{l_{1}=0}^{1}(-1)^{l_{1}}\left( 1+\sum_{j=2}^{n}(l_{1}-l_{j})\frac{%
i\sin \pi \nu }{\sinh \theta _{1j}}\right) e^{\theta _{1}-\frac{i\pi }{2}%
(1-(-1)^{l_{1}}\nu )} \\
\approx -ie^{\theta _{1}}\left( e^{\frac{i\pi }{2}\nu }-e^{-\frac{i\pi }{2}%
\nu }\right) +2i\sin \pi \nu
\,\sum_{j=2}^{n}\sum_{l_{1}=0}^{1}(-1)^{l_{1}}(l_{1}-l_{j})e^{\theta _{j}-%
\frac{i\pi }{2}(1-(-1)^{l_{1}}\nu )}
\end{multline*}
The first term again vanishes due to the lemma and the second one yields 
\begin{align*}
& 2i\sin \pi \nu
\,\sum_{j=2}^{n}\sum_{l_{1}=0}^{1}(-1)^{l_{1}}(l_{1}-l_{j})e^{\theta _{j}-%
\frac{i\pi }{2}(1-(-1)^{l_{1}}\nu )} \\
& =2i\sin \pi \nu
\,\sum_{j=2}^{n}\sum_{l_{1}=0}^{1}(-1)^{l_{1}}(l_{1}-l_{j})e^{\theta _{j}-%
\frac{i\pi }{2}(1-(-1)^{1-l_{j}}\nu )} \\
& =-2i\sin \pi \nu \,\sum_{j=2}^{n}e^{\theta _{j}-\frac{i\pi }{2}%
(1+(-1)^{l_{j}}\nu )}
\end{align*}
Therefore we finally obtain the asymptotic behavior 
\begin{eqnarray*}
\tilde{K}_{n}^{+}(\underline{\theta }) &=&-2i\sin \pi \nu
\sum_{l_{2}=0}^{1}\dots \sum_{l_{n}=0}^{1}(-1)^{l_{2}+\dots +l_{n}} \\
&&\times \prod_{2\leq i<j\leq n}\left( 1+(l_{i}-l_{j})\frac{i\sin \pi \nu }{%
\sinh \theta _{ij}}\right) \\
&&\times \sum_{i=2}^{n}e^{-\theta _{i}}\sum_{j=2}^{n}e^{\theta _{j}-\frac{%
i\pi }{2}(1+(-1)^{l_{j}}\nu )}+O(e^{-\mathop{\rm Re}\theta _{1}}) \\
&=&2i\sin \pi \nu \tilde{K}_{n-1}^{+}(\underline{\theta ^{\prime }})+O(e^{-%
\mathop{\rm Re}\theta _{1}})
\end{eqnarray*}
with $\underline{\theta ^{\prime }}=(\theta _{2},\dots ,\theta _{n})$. We
have used 
\begin{multline*}
\sum_{l_{2}=0}^{1}\dots \sum_{l_{n}=0}^{1}(-1)^{l_{2}+\dots
+l_{n}}\prod_{2\leq i<j\leq n}\left( 1+(l_{i}-l_{j})\frac{i\sin \pi \nu }{%
\sinh \theta _{ij}}\right) \\
\times \sum_{j=2}^{n}\left( e^{\theta _{j}-\frac{i\pi }{2}(1+(-1)^{l_{j}}\nu
)}+e^{\theta _{j}-\frac{i\pi }{2}(1-(-1)^{l_{j}}\nu )}\right) =0\,
\end{multline*}
which follows from lemma \ref{l1}. For $\tilde{K}_{n}^{-}(\underline{\theta }%
)$ and $\mathop{\rm Re}\theta _{1}\rightarrow \infty $ we have 
\begin{align*}
\sum_{i=1}^{n}e^{\theta _{i}}\sum_{i=1}^{n}e^{-z_{i}^{(l_{i})}}& =\left(
e^{\theta _{1}}+\sum_{i=2}^{n}e^{\theta _{i}}\right) \left( e^{-\theta _{1}+%
\frac{i\pi }{2}(1-(-1)^{l_{1}}\nu
)}+\sum_{i=2}^{n}e^{-z_{i}^{(l_{i})}}\right) \\
& =e^{\frac{i\pi }{2}(1-(-1)^{l_{1}}\nu )}+e^{\theta
_{1}}\sum_{i=2}^{n}e^{-\theta _{j}-\frac{i\pi }{2}(1-(-1)^{l_{j}}\nu )} \\
& ~~~+\sum_{i=2}^{n}e^{\theta _{i}}\sum_{i=2}^{n}e^{-z_{i}^{(l_{i})}}+O(e^{-%
\mathop{\rm Re}\theta _{1}}).
\end{align*}
We again calculate the leading terms $O(1).$ Again the contribution of the
first term consists of two types: one vanishes because of the lemma above
and the other is of order $O(e^{-\mathop{\rm Re}\theta _{1}})$. The
contribution of the third term vanishes after summation over $l_{1}.$ The
contribution of the second term is proportional to 
\begin{align*}
\sum_{l_{1}=0}^{1}& (-1)^{l_{1}}\left( 1+\sum_{j=2}^{n}(l_{1}-l_{i})\frac{%
i\sin \pi \nu }{\sinh \theta _{1i}}\right) e^{\theta
_{1}}\sum_{j=2}^{n}e^{-\theta _{j}+\frac{i\pi }{2}(1-(-1)^{l_{j}}\nu )} \\
& \approx 2i\sin \pi \nu \,\sum_{j=2}^{n}e^{\theta
_{i}}\sum_{l_{1}=0}^{1}(-1)^{l_{1}}(l_{1}-l_{j})e^{-\theta _{j}+\frac{i\pi }{%
2}(1-(-1)^{l_{j}}\nu )} \\
& =-2i\sin \pi \nu \,\,\sum_{j=2}^{n}e^{\theta _{i}}\sum_{j=2}^{n}e^{-\theta
_{j}+\frac{i\pi }{2}(1-(-1)^{l_{j}}\nu )}
\end{align*}
Therefore we finally obtain the asymptotic behavior 
\begin{eqnarray*}
\tilde{K}_{n}^{-}(\underline{\theta }) &=&-2i\sin \pi \nu
\sum_{l_{2}=0}^{1}\dots \sum_{l_{n}=0}^{1}(-1)^{l_{2}+\dots +l_{n}} \\
&&\times \prod_{2\leq i<j\leq n}^{n}\left( 1+(l_{i}-l_{j})\frac{i\sin \pi
\nu }{\sinh \theta _{ij}}\right) \\
&&\times \sum_{i=2}^{n}e^{\theta _{i}}\sum_{j=2}^{n}e^{-\theta _{j}+\frac{%
i\pi }{2}(1+(-1)^{l_{j}}\nu )}+O(e^{-\mathop{\rm Re}\theta _{1}}) \\
&=&-2i\sin \pi \nu \tilde{K}_{n-1}^{-}(\underline{\theta ^{\prime }})+O(e^{-%
\mathop{\rm Re}\theta _{1}})\,.
\end{eqnarray*}

\endproof %

Analogously one may discuss the behavior for $\mathop{\rm Re}\theta
_{1}\rightarrow -\infty .$

\end{document}